\documentclass[aps,prx,twocolumn,natbib,showpacs,floatfix,longbibliography,superscriptaddress]{revtex4-1}

\usepackage[unicode=true,
 bookmarks=false,
 breaklinks=true,pdfborder={0 0 1},backref=false,colorlinks=true]
 {hyperref}

\usepackage{graphicx}
\usepackage{amsmath,amssymb,bbold,bm,color}
\usepackage{float}
\usepackage{epstopdf}
\usepackage{hyperref}
\usepackage{color}
\usepackage{enumerate}
\usepackage{url}
\usepackage{braket}
\hypersetup{colorlinks=true,linkcolor=blue,citecolor=blue,urlcolor=blue}
\graphicspath{{figures/}{../figures/}}

\newcommand{\bk}{{\bm k}}

\newcommand{\br}{{\bm r}}
\newcommand{\bv}{{\bm v}}

\newcommand{\bmm}{{\bm m}}
\newcommand{\bA}{{\bm A}}

\newcommand{\bj}{{\bm j}}
\newcommand{\bsig}{{\bm \sigma}}

\newcommand{\bdel}{{\bm \delta}}

\newcommand{\cT}{{\cal T}}

\newcommand{\cP}{{\cal P}}
\newcommand{\cH}{{\cal H}}

\newcommand{\bee}{\begin{equation}}
\newcommand{\ee}{\end{equation}}

\begin{document}

\title{Persistent spin currents in superconducting altermagnets}

\author{Kyle Monkman}
\affiliation{Department of Physics and Astronomy, and Quantum Matter
  Institute, University of British Columbia, Vancouver, BC, Canada V6T 1Z1}

\author{Joan Weng}
\affiliation{Department of Physics and Astronomy, and Quantum Matter
  Institute, University of British Columbia, Vancouver, BC, Canada V6T 1Z1}

\author{Niclas Heinsdorf}
\affiliation{Department of Physics and Astronomy, and Quantum Matter
  Institute, University of British Columbia, Vancouver, BC, Canada V6T 1Z1}
\affiliation{Max Planck Institute for Solid State Research, Heisenbergstrasse 1, 70569 Stuttgart, Germany} 

\author{Alberto Nocera}
\affiliation{Department of Physics and Astronomy, and Quantum Matter
  Institute, University of British Columbia, Vancouver, BC, Canada V6T 1Z1}

\author{Marcel Franz}
\affiliation{Department of Physics and Astronomy, and Quantum Matter
  Institute, University of British Columbia, Vancouver, BC, Canada V6T 1Z1}
\begin{abstract}
Superconductors are famously capable of supporting persistent
electrical currents, that is, currents that flow without any
measurable decay as long as the material is kept in the
superconducting state. We introduce here a class of materials --
superconducting altermagnets -- that can both generate and carry
persistent {\em spin} currents. This includes spin-polarized
electrical supercurrent as well as pure spin supercurrent that facilitates
spin transport in the absence of any charge transport. A key to this
remarkable property is the realization that the leading superconducting
instability of altermagnetic metals consists of
two independent condensates formed of spin-up and spin-down electrons. In the
non-relativistic limit the two condensates are decoupled and can thus
naturally support persistent currents with any spin polarization,
including pure spin supercurrents realized in the charge counterflow
regime. We describe a novel ``spin-current dynamo effect''  that can
be used to generate pure spin supercurrent in such systems by driving
a charge current along certain crystallographic directions. 
Away from the non-relativistic limit, when spin-orbit
interactions and magnetic disorder are present, we
find that the spin current generically  develops spatial oscillations
but, importantly, no dissipation or decay. This is in stark 
contrast to spin currents in normal diffusive metals which tend to
decay on relatively short lengthscales.  We illustrate the
above properties by performing model calculations relevant to two distinct
classes of altermagnets and various device geometries. 

\end{abstract}

\date{\today}
\maketitle

\section{Introduction}
The magic of supercurrent lies in the fact that it flows in the ground
state of the system. In some geometries, such as a
superconducting (SC) ring threaded by magnetic flux that is some fraction
of the SC flux quantum $\Phi_0=hc/2e$, the current cannot decay even
as a matter of principle, simply because it flows in the quantum state
with the lowest energy \cite{Broom1961,File1963}. As a more practical
matter the fact that persistent currents are not subject to
dissipation has been widely exploited in devices, e.g.\ to generate
magnetic fields for a broad range of  applications. Quantum coherence
of the SC condensate also forms a basis for an important class of qubits
and quantum sensors based on the ubiquitous Josephson effect.

Unlike electric charge, spin is generically not conserved in solids.
Transporting spin over long enough distances so that it can be
successfully manipulated and read out is one of the key requirements of
spintronics \cite{Fabian2004,Sinova2015}. In diffusive normal metals and semiconductors spin
lifetimes are typically quite short owing to the spin-orbit coupling (SOC) and spin-flip
scattering. The quest in the spintronic
community to remedy this situation has led to the advent of
superconducting spintronics \cite{Robinson2015,Eschrig2015} which
seeks to exploit dissipationless 
currents carried by superconductors to achieve long spin transport
lifetimes. This effort has produced a number of notable advances
\cite{Blamire2018,Norman2018,Blamire2019,Farkhad2020} and  has
culminated in the recent demonstration of a
tunable pure spin supercurrent in heterostructures composed of a
ferromagnet, spin-orbit coupled metal and a conventional
superconductor such as Pt/Co/Pt/Nb/Ni$_8$Fe$_2$
\cite{Robinson2020}. Additionally, a number of theoretical proposals have
been advanced aimed at improving performance and attaining new
functionalities in superconducting spintronic devices
\cite{Nagaosa2019,Silaev2020,Montiel2023,Kamra2023,Bobkova2024}.

Despite this progress, platforms that
combine superconductors with ferromagnets have some inescapable drawbacks: Ferromagnets produce stray fields that can lead to undesirable coupling between distant circuit elements.  Also, conventional superconducting order
tends to be strongly suppressed by the magnetic field through both
orbital and Zeeman depairing. Likewise, magnetic exchange coupling is
pair-breaking for conventional spin-singlet Cooper pairs.  As a
result, it would be vastly preferable if one could generate and sustain spin-polarized
persistent currents in structures with zero net
magnetization. In what follows, we discuss one such platform based
on the recently discovered family of altermagnetic metals
\cite{ahn2019antiferromagnetism,Hayami2020,Smejkal2022a,Smejkal2022b,Mazin2022,Jiang2024}. Similar to
antiferromagnets, altermagnets exhibit zero net magnetization enforced
by their spin group symmetry \cite{ssgs1,ssgs2,ssgs3,ssgs4,ssgs5,ssgs6}. Nevertheless they possess spin-split fermi surfaces which,
as we will demonstrate, are naturally suited to form a basis for
exotic equal-spin triplet SC order capable of both generating and
supporting persistent spin currents.  

\begin{figure}[t]
  \includegraphics[width=8.5cm]{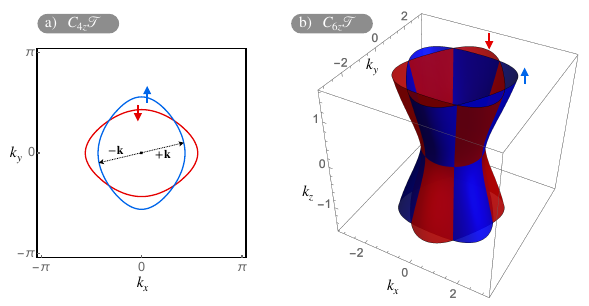}
  \caption{ Spin split Fermi surfaces of (a) $d$-wave and (b)
    $g$-wave altermagnetic metals 
    described by Hamiltonians Eq.\ \eqref{h1} and \eqref{h3} for altermagnetic band spitting parameter
    $\eta=0.2$. Blue and red color distinguishes up and down spin projections.}
  \label{fig0}
\end{figure}

The reason for this is quite simple and can be readily understood from Fig.\
\ref{fig0} which depicts spin-split fermi surfaces typical of two
families of altermagnets with 4-fold and 6-fold rotation symmetry,
also known as $d$- and $g$-wave altermagnets. We observe that these
types of fermi surfaces are fundamentally incompatible with the
conventional spin-singlet 
superconducting instability: for a spin-up electron at crystal
momentum $+\bk$ there is no requisite spin-down electron available at
$-\bk$. Instead, a zero-momentum Cooper pair must form from two electrons with equal
spin, which also necessitates odd-parity spatial symmetry of the pair
wave function $g_\bk=-g_{-\bk}$. This intuition has recently been
formalized by a general group-theoretic classification of symmetry-allowed SC phases
in altermagnets \cite{parshukov2025} which highlighted spin-triplet
order parameters as most natural SC instabilities.

Importantly for the task at hand, in the strictly
non-relativistic limit (that is, in the absence of SOC) the two spin
species are decoupled and so are the two superconducting
condensates. It is this feature that ultimately enables all the
interesting spintronic properties of such condensates discussed
below. For example, one could imagine setting up superflow in spin-up
condensate only: the resulting persistent current would then be fully
spin-polarized. Likewise, a state with the two condensates flowing in
the opposite direction realizes a pure spin supercurrent,
transporting spin with no concomitant charge transport.

In normal metals, inclusion of SOC and
spin-flip scattering is known to lead to short spin lifetimes, which are often the
main limiting factors in spintronics devices \cite{Fabian2004}. An important question
therefore arises as to the effect of such perturbations on the efficacy of the
proposed new platform. To address this question we perform an
extensive analysis in the framework of microscopic models as well
phenomenological Ginzburg-Landau (GL) theory
and show that SOC does not destroy these behaviors unless its
strength exceeds a critical value.  Specifically, we find that
inclusion of SOC causes spatial oscillations in spin supercurrent
amplitude along the flow direction, but, crucially, no decay or
dissipation. Hence, superconducting altermagnets can carry persistent spin currents
over long distances even when SOC is present. We also explore the
effect of spin-flip scattering caused by magnetic disorder and find
that unless it is very strong (so as to significantly suppress the SC order)
it also does not substantially degrade the spin supercurrent.

As argued previously
\cite{Zhu2023,Sudbo2023,Heung2024,Carvalho2024,Leraand2025} the
leading SC instability of a $d$-wave 
altermagnet in the presence of a weak attractive interaction occurs in
the equal-spin triplet channel with a fully gapped $p_x\pm ip_y$ order
parameter. This is relevant to altermagnets
with the square-lattice symmetry including the recently discovered
quasi-2D oxychalcogenide metals  KV$_2$Se$_2$O~\cite{Jiang2024} and
Rb$_{1-\delta}$V$_{2}$Te$_{2}$O~\cite{Zhang2024}. 
In the following we perform the same type of analysis for
$g$-wave altermagnets and show that the leading SC instability also
occurs in the equal-spin triplet $p_x\pm ip_y$ channel. This is potentially
relevant to a number of confirmed altermagnets with the hexagonal symmetry
including MnTe \cite{Krempasky2024,Lee2024}, RuO$_2$
\cite{Fedchenko2024} and CrSb \cite{Reimers2024,Ding2024,Yang2025}.

While it is true that none of the above altermagnets have been yet
reported to show superconductivity, research in this area is still in
very early stages and many more candidate altermagnet families have been 
identified by the recent theoretical work, both in naturally occurring crystals
\cite{Naka2019,Yuan2020,Mazin2021,Spaldin2022,Guo2023,Ostanin2024} and
in artificially engineered structures \cite{Liu2024b}. Since
many of these are good metals it would be surprising if at least some
did not show superconducting instabilities when cooled down to low
temperatures. It is also worth emphasizing that the predicted
unconventional equal-spin pairing does not require any exotic pairing
mechanism -- it emerges naturally as the leading SC instability of the
altermagnetic normal metal in the presence of conventional weak attraction mediated by
phonons \cite{Leraand2025} or any other bosonic mode.   

We remark that {\em proximity-induced} superconductivity in altermagnets has
been studied by a number of authors \cite{Linder2023,Papaj2023,Beenakker2023,Brataas2024,Neupert2024,Banerjee2024}. Structures formed of
conventional superconductors and altermagnets have been predicted to show a multitude of
interesting and potentially useful behaviors including
supercurrent-induced edge magnetization, Cooper pair spin-splitter and
filtering effect, controllable $0-\pi$ transitions in the Josephson
supercurrent, spin-polarized Andreev levels and Josephson diode
effect. The focus of this work, by contrast, is {\em intrinsic}
SC order  in altermagnets which remains relatively unexplored.


\section{Microscopic models}
\label{sec:model}

\subsection{$d$-wave altermagnet}

We consider a minimal model of a `$d$-wave' altermagnetic
metal \cite{Smejkal2022a,Smejkal2022b}  in two dimensions defined by the Hamiltonian
$\cH_0=\sum_\bk\psi_\bk^\dagger h_0(\bk)\psi_\bk$ with
$\psi_\bk=(c_{\bk\uparrow}, c_{\bk\downarrow})^T$ and
\begin{equation}\label{h1}
h_0(\bk)=-2t(\cos{k_x}+\cos{k_y})-2\eta\sigma_z(\cos{k_x}-\cos{k_y}).
\end{equation}
Here $\sigma_\mu$ are Pauli matrices in spin space, $t$ denotes the nearest
neighbor hopping amplitude on the square lattice. The $\eta$ term gives
the altermagnetic band spitting and hence breaks time-reversal
symmetry $\cT$. The model however remains invariant under combined $C_4$
rotation and $\cT$, a hallmark feature of altermagnets which
guarantees vanishing total magnetization. Unlike conventional antiferromagnetism, altermagnetic order does not enlarge the non-magnetic unit cell. Thus, the crystallographic basis of altermagnets is, in principle, required to contain at least two sites. While interband effects have been shown to be an important factor in the microscopic formation of altermagnetic order \cite{sitedecoupling1,sitedecoupling2}, the corresponding N\'eel temperatures tend to be orders of magnitude larger than the superconducting energy scales considered here. Consequently, we can safely neglect coupling between different sublattices and consider an effective single-band model \eqref{h1} with altermagnetic splitting introduced through spin-dependent tunneling amplitudes.  

When the inversion symmetry of $\cH_0$ is broken (e.g.\ by placing the
system on a substrate) then a Rashba SOC term
\begin{equation}\label{h2}
h_R(\bk)=2\lambda_R(\sigma_x\sin{k_y}-\sigma_y\sin{k_x})
\end{equation}
becomes symmetry-allowed. The main effect of non-zero $\lambda_R$ 
is to resolve the remaining  Fermi surface degeneracies along the
Brillouin zone diagonals visible in Fig.\ \ref{fig0}(a).
In the vicinity of these points  the Rashba term causes the spins to
rotate into the plane.

\subsection{$g$-wave altermagnet}

Although a purely 2D model of a $g$-wave altermagnet can be defined
\cite{Guo2025}, it requires a rather large unit cell which does not
correspond to any known naturally occurring material. We therefore take
inspiration from the crystal structure of CrSb, which is a confirmed
altermagnetic metal \cite{Reimers2024,Ding2024} and introduce an effective 3D
model on a layered triangular lattice which produces a qualitatively
correct spin-split fermi surface. The Hamiltonian is given as
\begin{eqnarray}\label{h3}
  h_0(\bk)=&-&2t\sum_{j}\cos{\bdel_j\cdot\bk}-2t_z\cos{k_z}
            \\
  &-&2\eta\sum_{j}\cos{[\sigma_zk_z+(\bdel_j-\bdel_{j-1})\cdot\bk]},  \nonumber 
\end{eqnarray}
where $\bdel_1=\hat{x}$ and $\bdel_{2,3}=-{1\over 2}(\hat{x}\mp
\sqrt{3}\hat{y})$ are nearest neighbor vectors on the triangular
lattice. $t_z$ denotes the interlayer tunneling amplitude and the
$\eta$ term produces the characteristic $g$-wave altermagnetic spin
splitting depicted in Fig.\ \ref{fig0}. This term breaks $\cT$ but the model
remains invariant under combined $C_{6z}$ rotation and $\cT$
which qualifies it as a $g$-wave altermagnet.

To account for the relativistic effects a Dresselhaus-type SOC of the
form 
\begin{equation}\label{h4}
  h_D(\bk)=2\lambda_D\sum_{j,\alpha=\pm}\sigma^{(j)}
  \sin{[\alpha k_z+(\bdel_j-\bdel_{j-1})\cdot \bk]},
\end{equation}
where $\sigma^{(j)}=\bsig\cdot[\hat{z}\times(\bdel_j-\bdel_{j-1})]$,
can be added. This term respects $\cT$ but is odd under inversion
$\cP$. Similar to $h_R$ its chief effect is to split the remaining
spin degeneracies located on high-symmetry 
planes that follow from the non-relativistic Hamiltonian Eq.\
\eqref{h3}. In addition, SOC terms Eq.\ \eqref{h2} and \eqref{h4}
break the $U(1)_\uparrow\times U(1)_\downarrow$ symmetry associated
with the separate number conservation of spin-up and spin-down
electrons down to a single $U(1)$ charge conservation symmetry. 

In the following, we will consider superconducting
instabilities of $\cH_0$ for both the $d$- and $g$-wave model
altermagnets in the presence of weak attractive interaction. When
specified we will study the effect of weak SOC with 
magnitude small compared to the SC gap, as well as magnetic disorder
described by a Hamiltonian specified below. We will take
$t=1$ and express all energies in units of $t$.

\subsection{Pairing instabilities}

To study SC instabilities of the altermagnetic normal metals
introduced above we consider Hamiltonian $\cH=\cH_0+\cH_I$ where
\begin{equation}\label{h5}
  \cH_0=\sum_\bk\psi_\bk^\dagger [h_\bk-\mu]\psi_\bk
\end{equation}
describes the underlying normal metal with $h_\bk$ defined as a sum of
$h_0(\bk)$ and the corresponding SOC terms given in Eqs.\
(\ref{h1}-\ref{h4}) above. For both $d$-
and $g$-wave cases we consider in-plane nearest-neighbor attraction
defined in real space as
\begin{equation}\label{h6}
  \cH_I=-V_1\sum_{\langle i,j\rangle}n_in_j
\end{equation}
where $V_1$ is a positive interaction parameter and $n_j=\sum_\sigma
c^\dagger_{j \sigma}c_{j \sigma}$ denotes the electron number operator
on site $j$.

We next perform standard mean-field decoupling of $\cH_I$ in
spin-singlet and equal-spin triplet channels and define the
corresponding SC order parameters 
\begin{equation}\label{h7}
\Delta_{ij}^0=V_1\langle c_{i\uparrow}c_{j\downarrow}\rangle, \ \ \ 
\Delta_{ij}^\sigma=V_1\langle c_{i\sigma}c_{j\sigma}\rangle, 
\end{equation}
associated with a bond between nearest neighbor sites $i$ and $j$. 
The singlet component is even under inversion,
$\Delta_{ij}^0=\Delta_{ji}^0$, and can be thought of as an extended
$s$-wave order parameter. The triplet component must be inversion-odd,
$\Delta_{ij}^\sigma=-\Delta_{ji}^\sigma$, and can be realized as a
$p$-wave order parameter.

Our discussion thus far has been completely general in that it applies to
both $d$- and $g$-wave altermagnets and, after recasting $h_0$ in real
space, the formalism can describe spatially non-uniform situations
such as in the presence of boundaries or currents. We now wish to study the
competition between singlet and triplet instabilities. To this end we
specialize to spatially uniform configurations of the order parameters
by assuming 
\begin{equation}\label{h8}
\Delta_{ij}^0=\Delta_0, \ \ \
\Delta_{ij}^\sigma=e^{\pm i\theta_{ij}}\Delta_\sigma, 
\end{equation}
where $\theta_{ij}$ denotes the angle between the bond and the
positive $x$ axis. $\Delta_{ij}^\sigma$ in Eq.\ \eqref{h8} describes the chiral $p$-wave order
parameter (with $\pm$ sign reflecting the two possible
chiralities). We expect such chiral $p_x\pm i p_y$ order parameter to
be energetically favored over pure $p_x$ or $p_y$ because it generally
leads to a fully gapped fermi surface.     In
the language of Ref.\ \cite{parshukov2025} these phases belong among
the family  ``spin chiral'' superconductors.   

In momentum space the decoupled interaction Hamiltonians take the
following form
\begin{subequations}
\begin{align}\label{h9a}
\cH_{\rm MF}^0=&&2\Delta_0\sum_\bk
                   C_\bk(c^\dagger_{\bk\uparrow}c^\dagger_{-\bk\downarrow}+{\rm
                   h.c.})+{4N\over V_1}\Delta_0^2,
  \\\label{h9b}
  \cH_{\rm MF}^\sigma=&&\Delta_\sigma\sum_\bk
                   (S_{\bk\sigma} c^\dagger_{\bk\sigma}c^\dagger_{-\bk\sigma}+{\rm
                   h.c.})+{2N\over V_1}\Delta_\sigma^2,
\end{align}
\end{subequations}
where on the square lattice 
\begin{equation}\label{h10}
                   C_\bk=\cos{k_x}+\cos{k_y}, \ \ \
                   S_{\bk\sigma}= \sin{k_x}\pm i\sin{k_y},
\end{equation}
and analogous expressions for the triangular lattice. This
leads to a description in terms of the  Bogoliubov-de Gennes (BdG) Hamiltonian 
\begin{equation}\label{h11}
\cH_{\rm BdG}=\sum_\bk\Psi^\dagger_\bk
\begin{pmatrix}
  h_\bk-\mu & \hat\Delta_\bk \\
 \hat\Delta_\bk^\dagger & -h_{-\bk}^\ast+\mu
\end{pmatrix} \Psi_\bk +E_0,
\end{equation}
where $\Psi_\bk=(c_{\bk\uparrow}, c_{\bk\downarrow},
c^\dagger_{-\bk\uparrow}, c^\dagger_{-\bk\downarrow} )^T$ is the Nambu
spinor,
\begin{equation}\label{h12}
\hat\Delta_\bk =
\begin{pmatrix}
\Delta_\uparrow S_{\bk\uparrow} & \Delta_0C_\bk\\
-\Delta_0C_\bk & \Delta_\downarrow S_{\bk\downarrow} 
\end{pmatrix},
\end{equation}
and $E_0=\sum_\bk{\rm Tr}(h_\bk)+{2N\over V_1}(2\Delta_0^2+\Delta_\uparrow^2+\Delta_\downarrow^2)$.

\begin{figure}[t]
  \includegraphics[width=8.6cm]{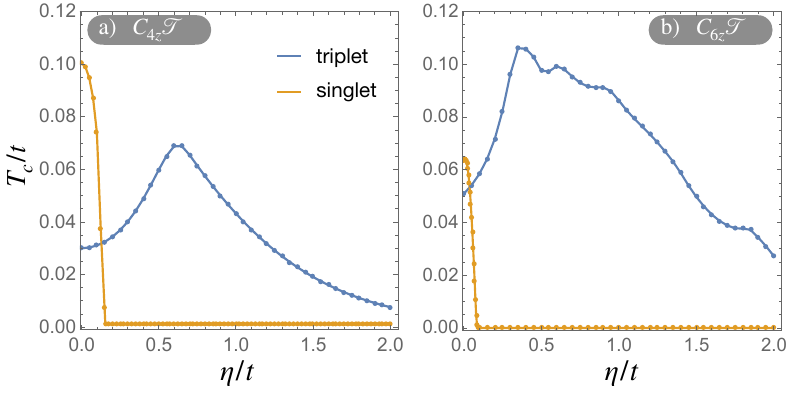}
  \caption{ Superconducting critical temperatures  in singlet and
    triplet channels extracted from
    numerical solutions to gap equations \eqref{h13a} and \eqref{h13b}
    for (a) $d$-wave and (b) $g$-wave model altermagnetic metals. We
    use $(\mu,V_1)=(-2.6,1.93)t$ for $d$-wave and $(-2.0,1.95)t$ for
    $g$-wave as representative parameters.
}
  \label{fig1}
\end{figure}
To study the competition between singlet and triplet channels we now
focus on the non-relativistic case where SOC terms are absent. The
system then respects the inversion symmetry and the two order
parameters belong to different irreducible representations of the
point group. In this situation, when only one order parameter is
non-zero, the $4\times 4$ BdG matrix in Eq.\ \eqref{h11} can be recast
as a $2\times 2$ matrix in the relevant subspace and can be easily
diagonalized. The corresponding BdG energy eigenvalues are found as
\begin{subequations}
\begin{align}
\label{h13a}
E_{\bk 0}&=&\xi_{\bk -}\pm\sqrt{\xi_{\bk +}^2+\Delta_0^2C_\bk^2}, \ \ \ \
  {\rm (singlet) } 
  \\
E_{\bk\sigma}&=&\pm\sqrt{\xi_{\bk\sigma}^2+\Delta_\sigma^2|S_{\bk\sigma}|^2}. \
                 \ \ \ \ \ \
  {\rm (triplet)} \label{h13b}
\end{align}
\end{subequations}
Here $\xi_{\bk\sigma}$ are the eigenvalues of $h_0(\bk)$ referenced to the
chemical potential $\mu$ and $\xi_{\bk\pm}=(\xi_{\bk\uparrow}\pm
\xi_{\bk\downarrow})/2$.

From the knowledge of the energy eigenvalues it
is straightforward to construct the free energy
$F$ of the system and obtain the
corresponding gap equations by minimizing $F$ with respect to the
order parameters. We thus find
\begin{subequations}
\begin{align}
\label{h14a}
\Delta_0&={V_1\over 2N}\sum_{\bk } {\Delta_0C_\bk^2\over \epsilon_{\bk}}{
\sinh{\beta \epsilon_{\bk}} \over \cosh{\beta \xi_{\bk -}} + \cosh{\beta \epsilon_{\bk }}}, 
  \\  \label{h14b}
\Delta_\sigma&={V_1\over 2N}\sum_{\bk } {\Delta_\sigma |S_{\bk\sigma}|^2\over E_{\bk\sigma}}
\tanh{{1\over 2}\beta E_{\bk\sigma}}, 
\end{align}
\end{subequations}
where $\epsilon_{\bk}=\sqrt{\xi_{\bk +}^2+\Delta_0^2C_\bk^2}$ and
$\beta=1/k_BT$ denotes the inverse temperature. We observe that 
while the gap equation for the triplet order parameter \eqref{h14b} 
assumes the standard BCS form, Eq.\ \eqref{h14a} is non-standard in
that it shows explicit dependence on the altermagnetic spin splitting
through $\xi_{\bk -}\sim \eta$. This reflects the expectation that
spin-singlet pairing will be sensitive to the $\cT$ breaking present in the
altermagnet while triplet pairing will not.

We solve Eqs.\ \eqref{h13a} and \eqref{h13b} numerically and show the
behavior of superconducting critical temperatures for singlet $s$-wave
and triplet $p_x\pm ip_y$ order parameters as a function of spin splitting $\eta$ in
Fig.\ \ref{fig1}. The behavior supports the general argument made in
Sec.\ I. As expected, the leading SC instability for a spin-degenerate
Fermi surface at $\eta=0$ occurs in the singlet channel which is signaled by
higher singlet $T_c$. However, even small values of
$\eta$ cause rapid suppression of the singlet $T_c$ -- altermagnetic spin splitting
is strongly pair breaking for singlet Cooper pairs. By contrast the
triplet $T_c$ is largely unaffected and even grows with increasing $\eta$.
(This initial rise of
$T_c$ is attributed to the increase with $\eta$ of the Fermi surface
size and hence the density of states $N_F$. At a constant interaction strength
$V_1$ the standard BCS expression $T_c\propto e^{-1/N_FV_1}$ then implies
increasing $T_c$.)

Crucially,  when $\eta\gtrsim \Delta_0$ the equal-spin triplet $p_x\pm
ip_y$ channel becomes the leading SC instability. In altermagnetic metals
$\eta$ is typically of the order of hundreds of meV and can be as large
as $\sim 1$eV. Typical values of $\Delta_0$ will be $1-10$meV. This leads to a
firm conclusion that superconductivity in any altermagnet will
almost certainly be in the equal-spin triplet channel.

For the sake of completeness we also
investigated the gap equation for the $p_x$ and $p_y$ case, which assume the same form as  Eq.\ \eqref{h14b} with
$S_\bk=\sin{k_{x/y}}$. The corresponding $T_c$ is more than an order
of magnitude smaller compared to the chiral $p_x\pm ip_y$ case for the
range of model parameters that we investigated. Hence, these
$p$-wave channels are not competitive.
\setlength{\tabcolsep}{6pt} 
\begin{table*}[t]
\centering 
\caption{Four degenerate ground states of the equal-spin triplet chiral
  $p$-wave superconductor in the non-relativistic limit.}
\begin{tabular}{c c c c c}
\hline\hline
 Ground state &\  ~\ ~\  Abbreviation  ~\  ~\ &\  ~\ ~\ Chern number \  ~\ ~\   &\  ~\ ~\ Property ~\ ~ \\
\hline \hline
$|\uparrow\uparrow (p_x+ip_y)\rangle\otimes |\downarrow\downarrow
  (p_x+ip_y)\rangle$ &            $p_+^\uparrow\otimes p_+^\downarrow$
                                    &  $+2$ &  chiral  \\
  \hline
  $|\uparrow\uparrow (p_x-ip_y)\rangle\otimes |\downarrow\downarrow
  (p_x+ip_y)\rangle$ &            $p_-^\uparrow\otimes p_+^\downarrow$
                                    &  $0$ &  helical  \\
  \hline
  $|\uparrow\uparrow (p_x+ip_y)\rangle\otimes |\downarrow\downarrow
  (p_x-ip_y)\rangle$ &            $p_+^\uparrow\otimes p_-^\downarrow$
                                    &  $0$ &  helical  \\
  \hline
  $|\uparrow\uparrow (p_x-ip_y)\rangle\otimes |\downarrow\downarrow
  (p_x-ip_y)\rangle$ &            $p_-^\uparrow\otimes p_-^\downarrow$
                                    &  $-2$ &  chiral  \\

\hline\hline
\end{tabular}
\label{table1}
\end{table*}

We note, finally, that the behavior depicted in Fig.\ \ref{fig1} is quite
generic, in that spin-singlet order parameter is always strongly
suppressed by altermagnetic splitting whereas spin-triplet is not. The
detailed dependence of critical temperatures is sensitive to the model
details but the above basic feature is a robust property of
altermagnetic metals. One can, for instance, add an on-site attraction $V_0$
to $\cH_I$ defined in Eq.\ \eqref{h6}. This has the effect of boosting
$T_c$ of the singlet channel for small $\eta$. Nevertheless, the
singlet $T_c$ is still strongly suppressed by spin splitting and 
the triplet order robustly prevails when $\eta > \Delta_0$.     

In the rest of this paper we focus on the limit of moderate to large
spin splitting parameter $\eta$ where the ground state is equal-spin
triplet chiral $p$-wave superconductor. In the non-relativistic limit
such a ground state exhibits a four-fold degeneracy due to two
possible chiralities for each spin. Two of the ground states are helical and two are
chiral as listed in Table 1. The assignment of the Chern number
follows from the fact that a spinless $p_x\pm ip_y$ state has
Chern number $C=\pm 1$, respectively \cite{fu2008}. Correspondingly, when placed on
a geometry with open edges, such as a long strip, the 4 ground states
will show protected edge modes that are either chiral for $C=\pm 2$
states, or helical for $C=0$ states. Because the chiral edge modes
carry substantial electrical currents \cite{Kallin2016}, one expects, for small samples,
the helical states to be lower in energy when one includes the
inductive effects, that is, the energy cost of the magnetic field $B$
produced by the edge current. As we shall see helical edge states carry
an aggregate spin current which however produces no magnetic field;
hence, we expect helical states to be lower in energy than chiral
states even in the strictly non-relativistic limit.      

In the presence of various perturbations the ground state degeneracy
will be lifted even in large samples. For the $C_{2z}\cT$ model this has been investigated in
Refs.\ \cite{Zhu2023,Heung2024}. According to these works Rashba SOC
selects helical $p_-^\uparrow\otimes p_+^\downarrow$ as the unique
ground state. In-plane magnetic field on the other hand selects the
two chiral states which remain degenerate. For the sake of concreteness
we will in the following assume weak SOC and focus on the helical
$p_-^\uparrow\otimes p_+^\downarrow$ state in our calculations. However, we note that a majority of the results pertaining to spin currents
presented in the rest of this paper are insensitive to the choice of
the ground state.

\section{Spin-polarized supercurrent generation in $d$-wave
  altermagnets}

\subsection{General considerations}

As our first example of useful spintronic properties of SC
altermagnets we show that they can be used to generate persistent spin
currents. This property is most pronounced in $d$-wave altermagnets
with equal-spin triplet superconducting order and relies on the fact
that the underlying fermi surfaces possess low $C_{2z}$ symmetry for
each spin projection, cf.  Fig.\ \ref{fig0}(a). In this situation it is easy
to see that the superfluid density tensor $\hat{\rho}_\sigma$, which
governs the supercurrent response of the material according to
\begin{equation}\label{h15}
\bj_\sigma=\hat{\rho}_\sigma\left(\hbar\nabla\varphi_\sigma-{2e\over c}\bA\right),
\end{equation}
will be anisotropic with {\em opposite} anisotropies for the two
spin projections.
Here $\bj_\sigma$ is the supercurrent density carried by condensate
$\Delta_\sigma$ and $\varphi_\sigma$ is its phase. Because Cooper
pairs with either spin carry the same charge $2e$ they
couple identically to the vector potential $\bA$. However, phases
$\varphi_\sigma$ can in general be different.    

The $C_{2z}$ symmetry of the band dispersion for each spin constrains the form of
$\hat{\rho}_\sigma$ such that
\begin{equation}\label{h16}
  \hat{\rho}_\sigma=
  \begin{pmatrix}
\rho_0 & 0\\
0&\rho_0
\end{pmatrix}
+
\sigma 
  \begin{pmatrix}
\rho_\eta & 0\\
0&-\rho_\eta
\end{pmatrix},
\end{equation}
where $\rho_0$ denotes the conventional isotropic component  and
$\rho_\eta$ is the anisotropy related to the altermagnetic spin
splitting, which is, crucially, opposite for two spin projections
labeled by $\sigma=\pm$. The
form of the superfluid density tensor Eq.\ \eqref{h16} is clearly
compatible with the $C_{4z}\cT$ symmetry of a $d$-wave
altermagnet. Microscopically it follows from the fact that the
superfluid density tensor is proportional to the inverse effective
mass tensor,
$[\hat{\rho}_\sigma]_{\alpha\beta}\propto\partial^2\epsilon_{\bk\sigma}/\partial
k_\alpha\partial k_\beta|_{\bk=0}$. For completeness we review the
relevant arguments in Appendix A where we also find, for the $d$-wave
altermagnet model
defined in Section \ref{sec:model},
\begin{equation}\label{h17}
\rho_0\simeq n_e, \ \ \ \rho_\eta\simeq (\eta/t) n_e,
\end{equation}
where $n_e$ denotes the electron density.

To explore the consequences of Eq.\ \eqref{h16} let us first assume, for the
sake of simplicity, that the two condensates experience the same
superfluid velocities $\bv_\sigma=(\hbar\nabla\varphi_\sigma-{2e\over
  c}\bA)$. This occurs e.g.\ when the current is driven principally by
$\bA$ or when the phases $\varphi_\uparrow$ and
$\varphi_\downarrow$ are locked together energetically by SOC. (We
will explore the effects of unequal velocities in later Sections.)  In
this situation it is easy to deduce from Eqs.\ \eqref{h15} and
\eqref{h16} that the electrical current $\bj^e=\bj_\uparrow+\bj_\downarrow$
will be governed by $\rho_0$ while the spin current
$\bj^s=\bj_\uparrow-\bj_\downarrow$ by $\rho_\eta$, namely 
\begin{equation}\label{h18}
\bj^e=\rho_0\bv, \ \ \ \bj^s=\rho_\eta\tau^z\bv.
\end{equation}
Here $\bv$ is the common superfluid velocity of the two condensates
and $\tau^z$ is the third Pauli matrix. From Eq.\ \eqref{h18}  one can also express the spin
current in terms of the electrical current as
\begin{equation}\label{h19}
\bj^s=(\rho_\eta/\rho_0)\tau^z\bj^e.
\end{equation}

\begin{figure}[t]
  \includegraphics[width=8.5cm]{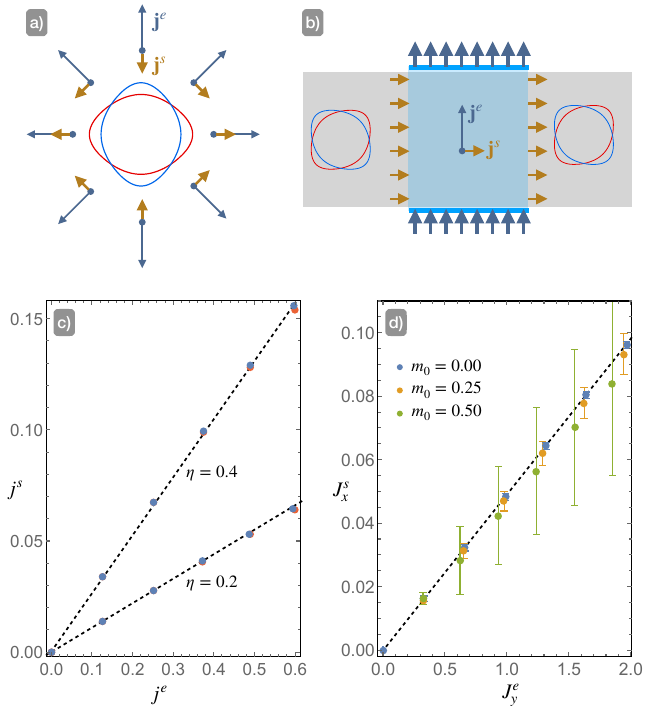}
  \caption{ Spin-polarized supercurrent generation in a $d$-wave
  altermagnet. Panel (a) visualizes the relation between $\bj^s$ and
  $\bj^e$ expressed by Eq.\ \eqref{h19}. Panel (b) illustrates the
  spin-current dynamo effect which may be used to generate pure spin
  supercurrent by applying ordinary charge current to a strip oriented
  along the  (1,1) direction relative to the underlying altermagnet
  principal axes. In (c) we show spin current density $j^s$ as a
  function of $j^e$ (both in the units of $t$) for a $20\times 20$
  cluster with periodic boundary conditions. Blue (orange) symbols
  correspond to $\bj^e$ flowing along $\hat{x}$ ($\hat{x}+\hat{y}$)
  direction. Panel (d) shows spin current $J^s_x$ as a function of
  $J^e_y$ in the dynamo geometry for the system studied in Fig.\
  \ref{fig3}. Data points represent the total spin current averaged
  over 40 positions along $\hat{x}$ while the error bar quantifies its
  fluctuation computed as a standard deviation of these values.  Parameters used are
  $\mu=-2.1$, $V_1=2.0$ with $\eta=0.2$ in panel (d).   
}
  \label{fig2}
\end{figure}

The simple-looking equation \eqref{h19}  has some interesting consequences
illustrated in Fig.\ \ref{fig2}(a). It implies that a persistent electrical
current $\bj^e$ directed along $x$ or $y$ principal axes will be
accompanied by a persistent 
spin current $j^s=(\rho_\eta/\rho_0)j^e$ along the same
direction. Even more remarkably, when $\bj^e$ is directed along the
diagonal, e.g.\ $(1,1)$ direction, spin current will flow along the
orthogonal $(1,-1)$ direction. As depicted in Fig.\ \ref{fig2}(b) this property can be
exploited to generate pure spin supercurrent with the magnitude
controllable by the applied charge current in the transverse direction. 
This pure spin supercurrent then can be  injected from the SC
altermagnet into another material; we call this {\em spin-current
  dynamo} effect and explore it in more detail below.
It can be thought of as a superconducting version of the
electrical spin splitter effect discussed in the context of organic
antiferromagnets and RuO$_2$ in Refs.\ \cite{Naka2019}
and \cite{Zelezny2021}, respectively. More broadly, the spin-current
dynamo effect belongs to the family of spin Hall effects that play an
important role in the field of spintronics \cite{Sinova2015}. 
%

\subsection{Spin current dynamo effect}

To confirm the validity of general arguments given in the previous
subsection  we now perform numerical
calculations using the BdG Hamiltonian Eq.\ \eqref{h11} formulated on
the real-space square lattice. We carry out two types of
calculations addressing different geometries. Specifically, we diagonalize the
real-space Hamiltonian for (i) a spatially uniform order parameter
$\Delta_\sigma$ on an $L\times L$ cluster with periodic boundary
conditions and (ii) on a $L_x\times L_y$ strip in the ``dynamo'' geometry
Fig.\ \ref{fig2}(b) with periodic boundary conditions along $x$ and
current injected through 10 sites at the bottom open edge and
extracted through top sites. In both
cases we probe currents by computing expectation values of the
spin-resolved  bond current operator
\begin{equation}\label{h18a}
j_{ij}^\sigma=-it_{ij}^\sigma
c^\dagger_{j\sigma}c_{i\sigma}+{\rm h.c.}
\end{equation}
where $t_{ij}^\sigma$ denotes the
tunneling amplitude between sites $i$ and $j$ for electrons with spin
$\sigma$.

Case (i) geometry corresponds to a torus and we drive
electrical current $\bj^e$ by inserting magnetic fluxes $\Phi_x$ and
$\Phi_y$ through the two holes of the torus This is implemented through
a Peierls substitution $t_{ij}^\sigma\to t_{ij}^\sigma
\exp{[-i(\br_i-\br_j)\cdot\bA}]$ where $\bA=\pi(\Phi_x,\Phi_y)/L\Phi_0$
is the vector potential. By adjusting the fluxes we
can drive charge current $\bj^e$  in any desirable
direction. Subsequent measurements of $\bj^s$ are summarized in Fig.\
\ref{fig2}(c) and  give results that are in
a good qualitative agreement with the prediction of Eq.\
\eqref{h19}. The relation between $j^s$ and $j^e$ is indeed linear with the
coefficient proportional to $\eta/t$ and about factor of 2 smaller
than the estimate based on the quadratic dispersion.  

\begin{figure*}[t]
  \includegraphics[width=17.5cm]{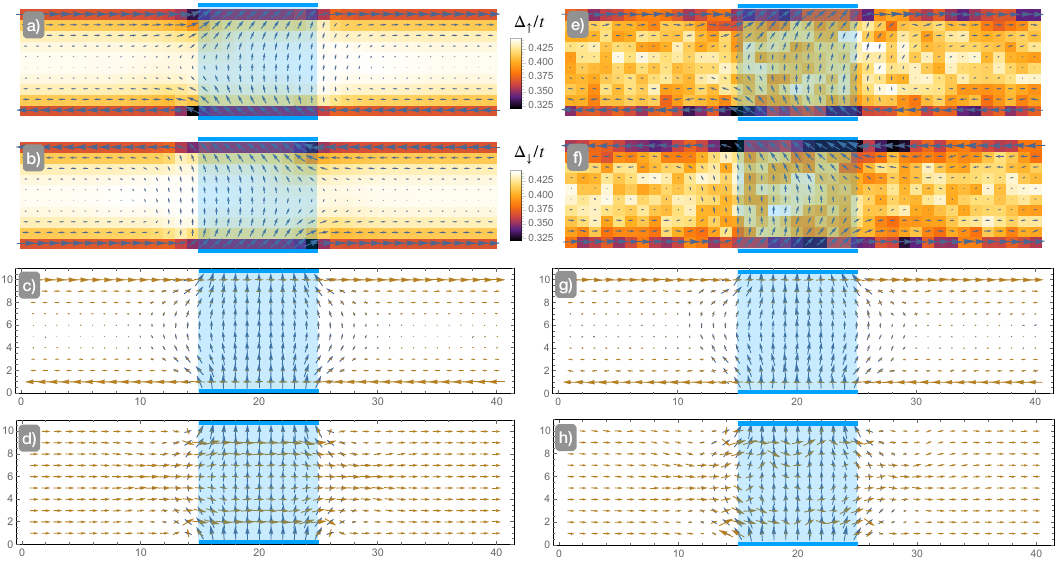}
  \caption{Spin-current dynamo effect: Generation of pure spin
    supercurrent from perpendicular charge current. Panels (a) and (b) show current densities
    $\bj_\uparrow$ and $\bj_\downarrow$ calculated from the BdG theory
    for a $40\times 10$ cluster of
    sites with periodic boundary conditions along $x$. The current is
    injected through the bottom blue electrode and collected from the top
  blue electrode. The colorscale indicates the amplitude of the
  self-consistently determined SC order parameter
  $\Delta_\sigma$. Panel (c) shows the current densities resolved into
the electrical component $\bj^e$ and spin component $\bj^s$,
represented by blue and orange arrows, respectively. Panel (d)
shows those same currents relative to the reference state when no
external current is injected. Panels (e-h) show the same quantities
for the system with magnetic disorder with $m_0=0.25t$. Spin currents
in panels (d,h) have been scaled up by a factor of 6 for better
visibility. Parameters are the same as in Fig.\ \ref{fig2} with
$t_J=1.0$ and $\Phi_x=0.4\Phi_0$. }
  \label{fig3}
\end{figure*}

Fig.\ \ref{fig3} shows our results for the dynamo geometry case (ii)
in a cluster of $40\times 10$ sites. In order to accommodate this
geometry naturally on a square lattice we modify the normal state
Hamiltonian by rotating the spin-dependent  term in Eq.\ \eqref{h1} by
$45^\circ$ which gives
\begin{equation}\label{h18b}
-2\eta\sigma_z[\cos{(k_x+k_y)}-\cos{(k_x-k_y})].
\end{equation}
This produces a $d_{xy}$ altermagnet with fermi surface crossings
along $x$ and $y$ axes and allows us to straightforwardly implement the
geometry indicated in Fig.\ \ref{fig2}(b). To probe the dynamo effect
we then drive charge current  in the vertical direction through the
central portion of the strip.  This is achieved by connecting sites at the bottom edge of
the strip with their partners at the same position $x$ at the top edge
through a spin-independent hopping amplitude
$t_Je^{-i\pi\Phi_x/\Phi_0}$. This creates a Josephson junction between a
portion of the bottom and top edge biased by flux
$\Phi_x$. When $\Phi_x$ is non-zero current flows through the
strip as indicated in Fig.\ \ref{fig3}(a,b) which shows the
corresponding spin-resolved current densities. We remark that in this
constrained geometry it is important to use self-consistently determined
order parameter amplitudes (shown as colorscale background) in order to
ensure current conservation.

Panel (c) displays current densities
resolved into their charge and spin components indicated by blue and
orange arrows, respectively. The spin current density
is dominated by prominent edge contributions flowing along the open edges of the
strip. These are indeed expected due to the presence of 
topologically protected edge modes in the
$p_-^\uparrow\otimes p_+^\downarrow$ superconductor which are known to
carry substantial currents \cite{Kallin2016}. Because of their
spin-filtered helical nature the corresponding charge current density vanishes
but significant spin current is seen to remain. The modes are however 
exactly  counterbalanced between the two edges and carry therefore no
net spin (or charge) current along 
the strip. In order to visualize the spin-current dynamo effect we
display in panel (d) spin current density referenced to the situation
with $\Phi_x=0$ where no charge current is driven across the strip. This subtraction
eliminates the edge currents and allows us to clearly see that  bulk spin current
is flowing along the strip.       

In order to ascertain the robustness of the spin current dynamo effect
we now consider the effect of magnetic disorder. To this end we
include a term
\begin{equation}\label{h18c}
\cH_{\rm dis}=\sum_j{\bmm_j}\cdot\bsig_{\alpha\beta}c^\dagger_{j\alpha}c_{j\beta}
\end{equation}
in the normal-state electron Hamiltonian  
where $\bmm_j$ represents a static local magnetic moment on site
$j$. We take Cartesian components of $\bmm_j$ as random independent variables
drawn from a uniform distribution between $(-m_0,m_0)$. This
corresponds to a random magnetic moment with mean-square amplitude
$\langle\bmm_j^2\rangle^{1/2}=m_0$ on every site of the lattice
coupled to electron spin.  Panels (e-h) of Fig.\ \ref{fig3} show the
effect of moderately strong disorder $m_0=0.25t$ which corresponds to
about half of the SC gap amplitude $\Delta_\sigma$. We observe clear
effects of disorder in local fluctuations of the gap amplitude,
although its average value is only mildly suppressed. The current flow
is also affected and shows visible local fluctuations in current
densities when compared to the clean case.

It is to be noted that spin current is no longer
conserved in the presence of $\cH_{\rm dis}$ because non-zero in-plane
components of $\bmm_j$ generate spin-flip scattering.
Nevertheless we observe in panel (h) a clear evidence of the dynamo
effect with a magnitude that is comparable to the clean case. This is
quantified in panel (d) of Fig.\ \ref{fig2}, where we show the
total spin current $J^s_x$ flowing along the strip in response to
charge current $J^e_y$ for three representative disorder strengths
$m_0$.  Because of the non-conservation mentioned
above $J^s_x$ fluctuates as a function of position $x$ along the
strip when $m_0\neq 0$. To provide a sense for these fluctuations  we display in Fig.\
\ref{fig2}(d) both the average value of $J^s_x$ and its standard
deviation. We observe that fluctuations are modest for $m_0=0.25$ but
become large for $m_0=0.50$, even though the average value is
suppressed only slightly. We remark that  $m_0=0.50$ corresponds to the
limit of very strong disorder with the magnetic impurity strength on each site
exceeding the size of the SC gap $\Delta_\sigma\simeq 0.42$. It is remarkable that
the spin current dynamo effect survives essentially intact even in this
highly disordered limit.

Some remarks are in order before we conclude this Section. First, we
note that the $C_{6z}\cT$ symmetry of $g$-wave altermagnets mandates
that $\rho_\eta=0$ and hence they cannot be used to generate spin
currents in this fashion; however, the effect might reemerge when strain is applied \cite{strain}. Interesting
properties besides spin current generation are explored below. Second, our general analysis leading
to Eq.\ \eqref{h19} relied
crucially on the simplifying assumption of equal superfluid velocities
$\bv_\sigma$ for the two spin species. In reality the two phases $\varphi_\sigma$ are
independent degrees of freedom that assume spatial profiles so as to minimize the system free
energy under the specific conditions imposed by the experimental
setup. Hence, generically, $\bv_\uparrow$ need not equal
$\bv_\downarrow$. Nevertheless for the geometries considered in this
Section we found Eq.\ \eqref{h19}  to be in a good agreement with the
numerical simulations in which  $\varphi_\sigma$ are treated as
independent variables and are computed selfconsistently from the gap equation. 

Finally, we note that the spin current generated by the dynamo
effect is odd under $\eta\to -\eta$. This implies that if multiple
magnetic domains are present in the sample their contributions to
$j^s$ will tend to average out. To achieve a significant effect one
thus ideally requires a single-domain sample or a sample dominated by
one magnetic domain. In the recent experiment on altermagnet MnTe domain
sizes greater than 10$\mu$m have been reported with careful annealing \cite{Amin2024}.

\section{Persistent spin current in various geometries}

A key conclusion reached in Section II was that the leading SC
instability in an altermagnet with large enough spin splitting
$\eta$ occurs in the equal-spin triplet channel. In the absence of SOC
this leads to two decoupled condensates, one for spin-up and one for
spin-down electrons represented by $\Delta_\uparrow$ and
$\Delta_\downarrow$ order parameters. At a high level it is
intuitively clear that such decoupled condensates are capable of
supporting persistent currents with an arbitrary spin polarization,
including pure spin supercurrent $\bj^s\neq 0$ with $\bj^e=0$,
achieved when spin-up charge current $\bj_\uparrow$ is exactly
counterbalanced by an opposite $\bj_\downarrow$.

To better understand various properties of these persistent currents
in different situations and also to understand the effect of SOC we now construct 
the  Ginzburg-Landau (GL) theory for the equal-spin triplet superconductor
 and analyze its solutions in various relevant geometries.

\subsection{Ginzburg-Landau theory}

The GL free energy density for the equal-spin triplet
superconductor can be written as
\begin{equation}\label{h20}
f[\psi_\uparrow,\psi_\downarrow]=\sum_\sigma
f_\sigma[\psi_\sigma]-{1\over 2}
D(\psi_\uparrow^\ast\psi_\downarrow+{\rm c.c.}) +{B^2\over 8\pi},
\end{equation}
where $\psi_\sigma$ are complex scalar order parameters describing the
two condensates and $f_\sigma$ is the standard GL free energy density
for the individual condensate,
\begin{equation}\label{h21}
f_\sigma[\psi]=\alpha|\psi|^2+{1\over
  2}\beta|\psi|^4+\gamma_\sigma|(i\nabla+{2e\over \hbar c}\bA)\psi|^2.
\end{equation}
By symmetry the $\alpha$ and $\beta$ parameters must be the same for
the two condensates. For a $d$-wave altermagnet the $\gamma$ parameter
becomes a tensor which reflects the superfluid density anisotropy
discussed in Sec.\ III. In the following, for the sake of maximum
clarity and simplicity, we focus on the isotropic case where  
$\gamma_\sigma$ is taken as a constant equal to $\gamma$ for both
condensates. Such theory is directly 
applicable to $g$-wave altermagnets and can also describe multidomain
$d$-wave altermagnets on long lengthscales where the anisotropy has
averaged out.

The $D$ term in Eq.\ \eqref{h20} describes coupling between the
two condensates and becomes symmetry-allowed in the presence of SOC
for the two helical states listed in Table 1.
In this situation we expect $D\propto \lambda^2$ where
$\lambda$ stands for either Rashba $\lambda_R$ in 2D or
Dresselhaus $\lambda_D$ in 3D. 
We note that for the two chiral states the $D$ term is not allowed.
As explained in Appendix B in this
case the leading allowed coupling is a fourth-order  term
$(\psi_\uparrow^{2\ast}\psi_\downarrow^2+{\rm c.c.})$ and has a
prefactor proportional to $\lambda^4$. 

Since we are mostly interested in the behavior of persistent currents on long
distances compared to the SC coherence length $\xi$ we adopt the London
approximation $\psi_\sigma=\psi_0e^{i\varphi_\sigma}$ with spatially
constant $\psi_0$, whose value we take as real and positive. This
leads to a Josephson free energy density of  the form  
\begin{equation}\label{h23}
f[\varphi_\uparrow,
\varphi_\downarrow]=f_0+\sum_\sigma\gamma\psi_0^2(\nabla\varphi_\sigma-{2e\over
  \hbar c}\bA)^2-J\cos{(\varphi_\uparrow-\varphi_\downarrow)}
\end{equation}
where $f_0$ is the component independent of the phases (which we
suppress from now on) and $J=\psi_0^2 D$. The last term  expresses the energy cost for configurations where two phases are not aligned and can therefore be thought of as describing an internal Josephson effect.  A direct calculation of $J$ from the microscopic $d$-wave
model is outlined in Appendix B  and gives
\begin{equation}\label{h22}
J\simeq
\lambda_R^2k_F^4/4\pi t,
\end{equation}
where $k_F=\sqrt{(\mu+4t)/t}$ is the dimensionless Fermi momentum. Eq.\
\eqref{h22} assumes the $p_-^\uparrow\otimes p_+^\downarrow$ helical phase and is valid for $\lambda_R\lesssim\Delta_\sigma$ and
$\mu$ near the bottom of the band, that is, $k_F\lesssim 1$.
\begin{figure*}[t]
  \includegraphics[width=17.5cm]{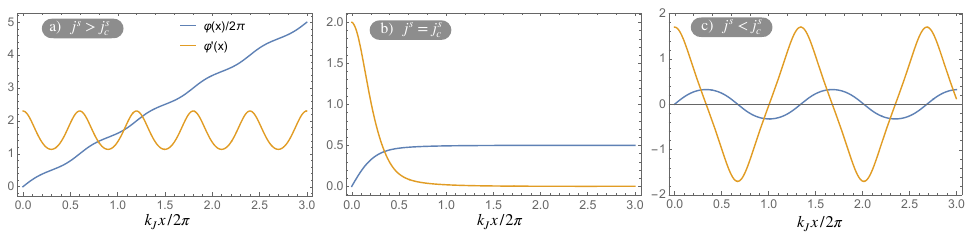}
  \caption{Evolution of a spin current $j^s(x)\propto \varphi'(x)$
    injected into a long wire at $x=0$ in the presence of SOC. The current is
    obtained by numerically solving Eq.\ \eqref{h28} with a boundary
    condition $\varphi(0)=0$ and $\varphi'(0)=(2.3, 2.0, 1.7)k_J$ in panels
    (a), (b) and (c), respectively. }
  \label{fig4}
\end{figure*}

It is convenient for our subsequent considerations to define average and
relative phases
\begin{equation}\nonumber
  \Omega={1\over 2}(\varphi_\uparrow+\varphi_\downarrow), \ \ \
  \varphi= \varphi_\uparrow-\varphi_\downarrow
\end{equation}
and rewrite Eq.\ \eqref{h23} as
\begin{equation}\label{h24}
f[\Omega,
\varphi]=\gamma\psi_0^2\left[2(\nabla\Omega-{2e\over
  \hbar c}\bA)^2+{1\over 2}(\nabla\varphi)^2\right]-J\cos{\varphi}.
\end{equation}
Charge current density then follows from varying the free energy $F=\int f dV$ with respect to $\bA$, which
gives
\begin{equation}\label{h25}
\bj^e=\bj_\uparrow+\bj_\downarrow=8e{\gamma\psi_0^2\over \hbar}(\nabla\Omega-{2e\over
  \hbar c}\bA).
\end{equation}
Similarly, we define the spin current density as
\begin{equation}\label{h26}
\bj^s=\bj_\uparrow-\bj_\downarrow=4e{\gamma\psi_0^2\over \hbar}\nabla\varphi.
\end{equation}

GL equations of motion follow from varying $F$ with respect to the two
phases. $\delta F/\delta \Omega =0$ gives $\nabla\cdot \bj^e=0$
expressing the electrical current conservation. $\delta F/\delta
\varphi =0$ yields an equation of motion
\begin{equation}\label{h27}
\gamma\psi_0^2\nabla^2\varphi +J\sin{\varphi}=0, 
\end{equation}
which must be solved subject to the appropriate boundary conditions in
order to find the spin current. Below we discuss some specific
examples.

\subsection{Spin-polarized supercurrent in a long wire}

A question of practical interest for any spintronic application is
this: If we inject spin-polarized current into a wire made of a SC
altermagnet, how far from the injection point will the spin
polarization persist? In the following we 
address this question by studying the GL theory derived above for a
long wire geometry with the 
spin-polarized current injected from the left. We assume a homogeneous
wire with the current injected at $x=0$ flowing along the positive
$x$ direction.  We consider a charge current with an  arbitrary spin polarization expressed
by specifying values of $j^e$ and $j^s$.

We first consider the non-relativistic limit. In this case there is no
SOC which translates to  $J=0$. Eq.\ \eqref{h27} can
then  be recast as $\nabla\cdot \bj^s=0$ expressing spin current
conservation exactly as one would expect  when the two condensates are
decoupled. Physically, this implies that {\em any} spin-polarization present
in the injected current will persist to arbitrarily long lengthscales.

Away from the non-relativistic limit we have to include the effect of non-vanishing
$J$. Qualitatively, it is clear from Eq.\ \eqref{h26} that nonzero spin
current requires nonzero gradient of the relative phase
$\varphi$. This also means that $\varphi$ cannot be uniformly zero to
minimize the $-J \cos{\varphi}$ term in the free energy Eq.\
\eqref{h24}:  the spin current will incur an energy cost. To see how
this impacts the spin current amplitude along the wire we recast Eq.\
\eqref{h27} as
\begin{equation}\label{h28}
{d^2\varphi\over d x^2} +k_J^2\sin{\varphi}=0, 
\end{equation}
where $k_J=\sqrt{J/\gamma\psi_0^2}$ is a constant with the dimension
of inverse length. This is a second-order differential equation which
we want to solve subject to boundary conditions at $x=0$ for $\varphi$
and its first derivative $\varphi'$. The latter is determined by the amplitude of
the injected spin current through Eq.\ \eqref{h26} while
$\varphi_0=\varphi(x=0)$ can be fixed
by physical conditions at the boundary, e.g.\ when the current is
injected from another superconductor.

To gain some intuition for the solutions of Eq.\ \eqref{h28}, it is
useful to note that it is mathematically equivalent to the equation of
motion for the angle of a mechanical pendulum,  
\begin{equation}\label{h29}
{d^2\varphi\over d t^2} +\omega^2\sin{\varphi}=0, 
\end{equation}
where $\omega=\sqrt{g/l}$ is the frequency of small oscillations. ($g$
and $l$ are the gravitational constant and the pendulum length,
respectively.) Based on this analogy an
important observation can be made right away. Because Eq.\ \eqref{h28}
corresponds to an {\em undamped} pendulum motion, we may conclude
immediately that the spin current injected into the wire will not decay;
instead, it will exhibit oscillations as a function of $x$ with a
constant amplitude. In analogy with the frictionless mechanical pendulum one expects two
regimes for these oscillations: (i) the current can oscillate around
zero which is analogous  to small oscillations of the pendulum about
the stable equilibrium at $\varphi=0$, or (ii) it can oscillate around a non-zero value
$j^s_{\rm avg}$ which corresponds to the pendulum rotating around the
pivot with a steadily increasing angle. The two regimes are separated
by a critical point where the pendulum has just enough energy to reach
the unstable equilibrium at the top of its trajectory.        

Solutions of Eq.\ \eqref{h28} displayed in Fig.\ \ref{fig4} indeed
confirm these expectations. When large enough spin current $j^s>j_c^s$
is injected panel (a) shows amplitude oscillations about a non-zero
average value $j^s_{\rm avg}$ which persist without any decay to
arbitrary distances. When smaller current $j^s<j_c^s$ is injected then
the amplitude still oscillates without any decay but now around zero,
panel (c). Only when the injected current is exactly tuned to the critical
value $j^s=j_c^s$ it decays, panel (b). We note, however, that because
this is an unstable fixed point (corresponding to a precisely tuned
total energy in the 
pendulum analogy) an oscillatory behavior of one or the other type
will generically occur. We conclude that even in the presence of SOC
spin-polarized current injected into the wire will propagate to
arbitrary distances without any decay. For $j^s<j_c^s$ one must detect
it away from the nodes of the oscillatory profile depicted in Fig.\
\ref{fig4}(c) whereas for $j^s>j_c^s$ it can be detected everywhere
along $x$.

To connect with experiment it is useful to estimate the period of
spin-current oscillations and the critical spin current $j_c^s$ which
separates the two regimes. The period, given by $2\pi/k_J$, follows
from recalling the well-known GL relations \cite{tinkham2004} for
$\psi_0^2=-\alpha/\beta$, the coherence length
$\xi^2=\gamma/|\alpha|^2$  and condensation energy ${1\over
  2}N_F\Delta_0^2=\alpha^2/2\beta$, where $N_F\simeq 1/4\pi t$ is the
density of states at the Fermi level. Combining these with Eq.\
\eqref{h22} one obtains
\begin{equation}\label{h30}
{2\pi\over k_J}\simeq 2\pi\xi\left({\Delta_0\over
    \lambda_R}\right)k_F^{-2}, 
\end{equation}
valid when $\lambda_R\lesssim \Delta_0$. We observe that the period
diverges when $\lambda_R\to 0$ and $\mu\to -4t$, in accord with
expectations: in these limits coupling $J$ between the two condensates
vanishes. Even away from these extreme limits Eq.\ \eqref{h30} predicts
the period of oscillations to exceed the GL coherence
length $\xi$, which justifies our usage of the London approximation to
derive these results.

The critical spin current density corresponds to $\varphi'(0)=2k_J$ which
together with Eq.\ \eqref{h26} gives
$j_c^s=8e(\gamma\psi_0^2/\hbar)k_J$. Using the relations listed above
Eq.\ \eqref{h30} one can express it more usefully as
\begin{equation}\label{h31}
j_c^s\simeq j_c^e\left({\lambda_R\over\Delta_0}\right)\sqrt{3}k_F^2, 
\end{equation}
where $j_c^e=8e(\psi_0^2/3\hbar)\sqrt{\gamma|\alpha|/3}$  is the GL
critical current density \cite{tinkham2004}.  We observe that $j_c^s$ will be small in
comparison with $j_c^e$ when either SOC is weak or when $\mu$ is near
the bottom of the band.

A natural question arises when pondering the results in Fig.\ \ref{fig4}, namely, to what extent is the oscillatory behavior dependent on the prefect translational invariance of the system. In other words, is this behavior stable with respect to disorder and inhomogeneity? We address this question in Appendix C and find that moderate levels of disorder, modeled as a random spatially varying component of the Josephson coupling $J$, do not substantially modify our conclusions. Specifically, the spin current injected into a long wire still propagates without dissipation; the main effect of disorder is to add a degree of irregularity to oscillations as illustrated in Fig.\ \ref{fig7}.    

\begin{figure}[t]
  \includegraphics[width=8.5cm]{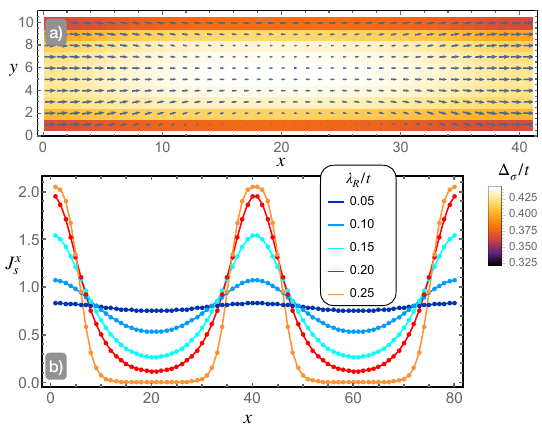}
  \caption{Effect of Rashba SOC on pure spin supercurrent in a ring
    geometry modeled as a  $80\times 10$ strip with periodic boundary conditions
    along $x$, open along $y$. (a) Order parameter amplitude
    $|\Delta_\sigma|$ (identical for both spins) shown as
   the  colorscale background and spin current density distribution
    $\bj_s$ indicated by arrows. Both quantities are shown for  the
    left half of the strip,  the right half is
    identical. In order to filter out the edge current we subtracted $\bj_s$ calculated for the same
    parameters with $n_v=0$.  (b) Total spin 
    current $J_s^x$ as a function of the position $x$ along the strip
    for several representative values of Rashba SOC. The same values of
    BdG model parameters are used as in Fig.\ \ref{fig2} with $n_v=1$
    and $\lambda_R=0.15$ in panel (a).}
  \label{fig5}
\end{figure}
To further support the above predictions of the GL theory we performed analogous
calculations within the microscopic model defined by the  BdG Hamiltonian
Eq.\ \eqref{h10} with Rashba SOC implemented via Eq.\ \eqref{h2}. We
employ here the same long strip geometry with $d$-wave nodes aligned
with the principal axes  as in Sec.\ III.B and with periodic
boundary conditions along $x$, open along $y$, forming topologically a
ring. To generate
spin current along $x$ we initialize the order parameters according to
\begin{equation}\label{g31}
\Delta_\sigma=\Delta_0e^{2\pi i\sigma n_v x/L_x}, 
\end{equation}
with $n_v$ denoting the integer vorticity, opposite for two spin
projections. For non-zero $n_v$ this choice describes 
spin-up and down condensates whose relative phase $\varphi$ winds by $4\pi
n_v$ around the ring. When
$\lambda_R=0$ the two condensates are decoupled and one trivially
obtains the charge  counterflow regime with uniform pure spin current 
circulating the ring. 

The effect of non-zero $\lambda_R$ on the spin current is illustrated in Fig.\
\ref{fig5} where we display fully self-consistent solutions of the BdG
theory starting from the ansatz Eq.\ \eqref{g31}. We remark that
achieving full self-consistency between the  BdG Hamiltonian
Eq.\ \eqref{h10}, suitably re-cast in the real space, and the corresponding gap equations \eqref{h7} is 
essential for obtaining physical, charge-current conserving solutions. We also
note that under this iterative procedure the ansatz Eq.\ \eqref{g31} leads to stable solutions
with non-zero phase winding, even though the global ground state is in
the zero-winding sector.  

In agreement with the results of the GL theory we find that
SOC brings about spatial oscillations in the spin current $J_s^x$ along the
ring, Fig.\ \ref{fig5}(b). The amplitude of oscillations increases with  $\lambda_R$, but,
importantly, there is no decay in the sense that solutions of the BdG
equations are always spatially periodic. This remains true even for very long
strips; we were able to simulate systems up to  $160\times 10$ in size
and various values of $n_v$, all with similar outcomes.

For larger values of $\lambda_R\gtrsim 0.22$ the spin current
is seen to vanish in segments of the 
strip. We associate this behavior with reaching and exceeding the
critical strength of SOC for a given value of the spin
current. According to the GL results in a semi-infinite geometry we expect in
this regime behaviors like those shown in Fig.\ \ref{fig4}(b) and (c),
respectively. In our present ring geometry the relative phase
$\varphi$ is constrained to wind by $4\pi n_v$ and hence behaviors of
this type cannot be reached. Instead, the system shows a transition
between the regime with non-zero $J_s^x$ everywhere along the ring to
the regime with $J_s^x=0$ along parts of the ring. Physically, the phases of two condensates are locked together in these regions and hence $J_s^x=0$. The relative phase varies rapidly elsewhere, resulting in large $J_s^x$, in accord with Eq.\ \eqref{h26}.

\subsection{Persistent spin current in a thin ring}

\begin{figure}[t]
  \includegraphics[width=8.1cm]{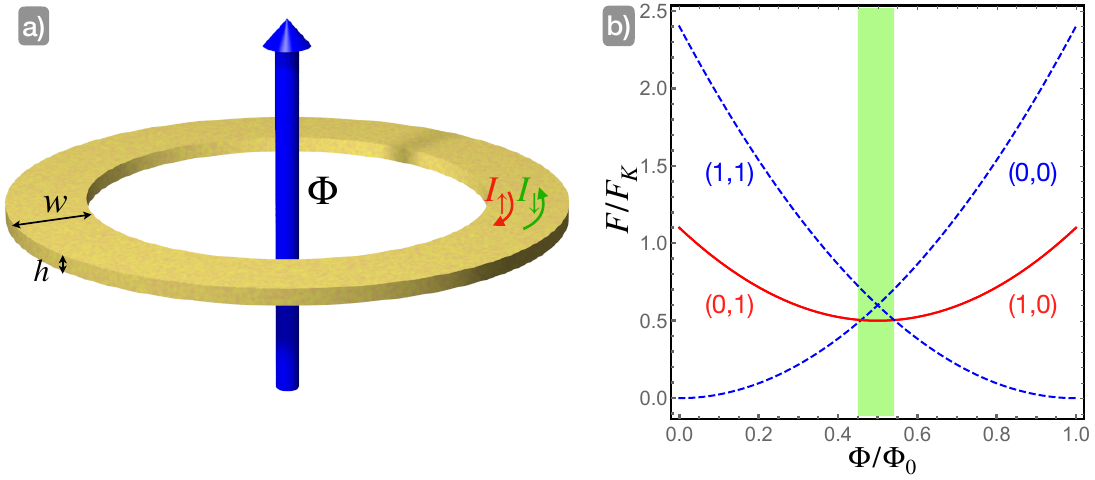}
  \caption{(a) A SC ring threaded by magnetic flux $\Phi$. Red and green
    arrows illustrate the charge counterflow regime that becomes
    energetically favorable when $\Phi$  is close to a half-integer
    multiple of the SC flux quantum at weak SOC. (b) Free energy Eq.\
    \eqref{h34} as a function the inserted flux $\Phi/\Phi_0$. Various
    branches correspond to winding numbers $(n_\uparrow,n_\downarrow)$
  as indicated and $F_K={1\over 2}L_KI_0^2$. The green-shaded region
  marks the range of flux values where charge counterflow solutions minimize the
  free energy and are thus stable.}
  \label{fig6}
\end{figure}

We now consider a ring threaded by magnetic flux $\Phi$, illustrated
in Fig.\ \ref{fig6}. We focus here on
the limit where the ring thickness $h$ and width $w$ are both smaller
than the magnetic penetration depth. This can be achieved if the ring
is made of a thin film such that $h\ll \lambda_L$, the bulk London
penetration depth. In this situation the effective in-plane penetration
depth will be given by the Pearl length, $\lambda_{\rm
  Pearl}=2\lambda_L^2/h\gg \lambda_L$, making the condition $w\ll \lambda_{\rm
  Pearl}$ easy to satisfy.

In the situation described above, the ring material will not quantize
the total enclosed flux and the calculations become straightforward as
we do not need to consider screening effects. In the non-relativistic
limit ($J=0$) the current densities contributed by the two decoupled
condensates follow from Eq.\ \eqref{h23} and read
\begin{equation}\label{h32}
\bj_\sigma=4e{\gamma\psi_0^2\over \hbar}(\nabla\varphi_\sigma-{2e\over
  \hbar c}\bA). 
\end{equation}
We represent the flux by  $\bA=\hat{\theta}\Phi/\ell$, where $\hat{\theta}$ is a unit
vector in the azimuthal direction and $\ell$ denotes the ring
circumference. Phases $\varphi_\sigma$ are constrained by
single-valuedness of the corresponding order parameters
$\psi_\sigma$. This permits the phase to wind by an integer multiple of $2\pi$ around the ring,
$\varphi_\sigma(\ell)=\varphi_\sigma(0)+2\pi n_\sigma$, where
$n_\sigma$ is the integer winding number and we
parametrized $\varphi_\sigma$ by its position along the ring. Assuming
furthermore that phases vary uniformly along the ring we have
$\nabla\varphi=2\pi n_\sigma \hat{\theta}/\ell$. Taken together this
allows us to define the ring current  
\begin{equation}\label{h33}
I_\sigma=S\hat{\theta} \cdot\bj_\sigma=I_0 (n_\sigma-\Phi/\Phi_0), 
\end{equation}
for each spin, where $I_0=4eS{\gamma\psi_0^2\over \hbar}{2\pi\over
  \ell}$ and  $S=wh$ is the ring cross section.

To understand the energetics (working still in the $J=0$ limit) it is useful to write the free energy
associated with these currents as
\begin{equation}\label{h34}
F={1\over 2}L_K(I_\uparrow^2+I_\downarrow^2)+{1\over
  2}L_G(I_\uparrow+I_\downarrow)^2. 
\end{equation}
The first term represents the energy cost of the superflow and follows
directly from the free energy density Eq.\ \eqref{h23} with the kinetic
inductance defined as $L_K=\hbar^2\ell/8e^2S\gamma\psi_0^2$. The
second term expresses the energy stored in the magnetic field $B$
generated by the total electric current $I= I_\uparrow+I_\downarrow$
flowing in the ring with $L_G$ denoting its geometric inductance and
follows from the last term in Eq.\ \eqref{h20}. The precise 
value of $L_G$ depends on the details of the ring geometry. For the illustration
purposes we can estimate it by the well-known Kirchhoff formula 
$L_G\simeq (8\pi\ell/c^2)[\ln(\ell/\rho)-1.508]$ for the self-inductance of a ring
with circular cross section of radius $\rho$ and circumference $\ell$,
accurate in the limit $\ell \gg\rho$ \cite{Cohen1907}.

A simple but important observation follows from Eq.\ \eqref{h34}: In
situations when the magnitudes of  $I_\uparrow$ and 
$I_\downarrow$ are equal, the free energy is minimized when
$I_\uparrow=-I_\downarrow$ because then the second term vanishes. This
situation 
corresponds to the {\em charge counterflow} regime in which  there is a pure
spin current flowing in the wire. The physical reason behind this interesting
conclusion is simply that while charge current incurs an energy cost
due to the induced magnetic field, spin current does not.

A practical route towards the realization of the charge counterflow
regime offers itself based on the above insight. Fig.\ \ref{fig6}(b)
shows the behavior of the free energy Eq.\ \eqref{h34} as a function
of the flux $\Phi$ with various branches representing different
vorticities $(n_\uparrow,n_\downarrow)$. In equilibrium the system
will be on the branch that minimizes $F$ for the given flux. We
observe that close to $\Phi=\Phi_0/2$ the global minimum occurs when
$(n_\uparrow,n_\downarrow)=(1,0)$ or $(0,1)$. In view of Eq.\
\eqref{h33} these solutions correspond to spin-up and down currents
flowing in the opposite directions. At exact half flux quantum this
implies pure spin persistent currents flowing in clockwise
and counterclockwise direction for $(n_\uparrow,n_\downarrow)=(1,0)$ and
$(0,1)$,  respectively. Solutions with winding numbers
$(1,1)$ and $(0,0)$ correspond to conventional (unpolarized)
persistent currents but are higher in energy by ${1\over 2}L_GI_0^2$.

We thus conclude that in the non-relativistic limit one can realize pure
spin supercurrent in a ring threaded by half-integer number of
magnetic flux quanta. The two states with this property minimize the global free
energy and are thus truly persistent in the sense that spin currents
cannot decay even as a matter of principle.

A natural question arises: How does a departure from the
non-relativistic limit affect this result? Persistent spin
currents discussed above involve solutions 
with different winding numbers $n_\sigma$ for the two
spin projections. Therefore, the phase difference
$\varphi_\uparrow-\varphi_\downarrow$ in the argument of the $J$ term
in the free energy Eq.\ \eqref{h23} will also wind around
the ring. This implies an energy cost $\Delta F_J\simeq JS\ell$
compared to the state with $\varphi_\uparrow=\varphi_\downarrow$. On
this basis we expect the persistent spin current solutions  to be energetically
favored as long as   $\Delta F_J\lesssim {1\over 2}L_GI_0^2$. This
places an upper bound on the strength of SOC that can be tolerated, expressed as
$J<J_c\simeq L_GI_0^2/2S\ell$. Using Eq.\ \eqref{h22} and various
relations listed above Eq.\ \eqref{h30} one can translate this into a
condition $\lambda_R<\lambda_{Rc}$ with
\begin{equation}\label{h36}
 \lambda_{Rc}^2\simeq {4\pi^2\Delta_0^2 \over
   k_F^4}\left({\xi\over\lambda_L}\right)^2{S\over \ell^2}\left[\ln{\ell\over\rho}-1.508\right].
\end{equation}
Here $\lambda_L^2=\hbar^2 c^2/8\pi(2e)^2\gamma\psi_0^2$ is the London
penetration depth and $\rho\simeq\sqrt{S/\pi}$ is the effective radius
of the ring cross section area.

We thus find that persistent spin current solutions can be favored even
away from the non-relativistic limit as long as the SOC strength satisfies
$\lambda_R<\lambda_{Rc}$ with $\lambda_{Rc}$ given in Eq.\
\eqref{h36}. To get a sense for the relevant magnitudes we consider a ring with radius $R=1\mu$m
and a rectangular 
cross section with $(w,h)=(0.1,0.01)\mu$m. If we furthermore assume
representative parameter values
$\xi/\lambda_L\approx 1$ and $k_F=0.5$ then Eq.\ \eqref{h36} gives
$\lambda_{Rc}\simeq 0.3\Delta_0$. Hence we conclude that under rather
generic conditions there will exist a reasonable window of permissible
SOC strength in which pure persistent spin current can be induced in a
ring by half-integer flux. This window, according to  Eq.\
\eqref{h36}, becomes wider for smaller samples, a useful property that
can be used detect and tune the effect experimentally. 

We note, finally, that the pure spin supercurrent predicted to occur near the half-flux quantum can be detected by a simple magnetization measurement, i.e.\ no direct spin-current detection is required. Consider increasing $\Phi$ from zero. Initially, the system will be on the (0,0) branch in Fig.\ \ref{fig6}(b) and will hence support charge current proportional to $\Phi$ according to Eq.\ \eqref{h33}, accompanied with the corresponding magnetization, also proportional to $\Phi$. Sufficiently close to $\Phi_0/2$ the system will switch to the (1,0) or (0,1) branch. Here the charge current is near zero and so is the magnetization. The signature of a pure spin supercurrent is therefore given by a range of near-vanishing magnetization around $\Phi=\Phi_0/2$ -- this is in contrast to the normal case where magnetization is maximal in this region.

\section{Outlook}

Efficient generation, manipulation, and detection of spin currents is one
of the key goals of the field of spintronics. Our work establishes
superconducting altermagnets as a new viable platform for achieving
some of these goals in a setting where spin dissipation, a key
challenge that must be faced in normal metal and semiconductor devices, is not an
issue. We have demonstrated that SC altermagnets are capable of
carrying spin-polarized currents without dissipation over arbitrarily
long distances, even in the presence of SOC and
magnetic disorder which generally lead to short spin life times in
non-superconducting materials.

The above advances are rooted in two key
insights: (i) the characteristic spin-split band structure of metallic altermagnets makes them
uniquely susceptible to an exotic form of spin-triplet $p$-wave
superconductivity, perfectly suited for generation and transport of spin
currents; and (ii) the resulting spin currents are persistent in that they flow in
the equilibrium state of the system, and hence, once established they
cannot decay.  This last property is well documented in the context of conventional
charge supercurrents. Here we showed that with small modifications it
extends to spin supercurrents.  We found that while perturbations away from the strictly non-relativistic limit (e.g.\ SOC and magnetic disorder) can cause 
spin current to undergo spatial oscillations and fluctuations,
crucially, it still propagates over arbitrary distances without any
overall decay. The difference is rooted in the fact that unlike charge, spin is not a conserved quantity in real materials and hence spin polarization is allowed to fluctuate. Nevertheless, the structure of the theory still does not permit dissipation to occur simply because the system with spin supercurrent is already in the state of the lowest free energy compatible with its boundary conditions.

A key challenge going forward lies in discovery of a suitable
family of altermagnets with an intrinsic superconducting instability occurring at temperatures that are not too low to prevent practical applications. Although no SC altermagnets have yet been identified, it is
to be noted that many existing and predicted altermagnets are good
metals. With zero net magnetization enforced by spin group symmetry
such metals are generically susceptible to Cooper pair formation in
the presence of a weak attractive interaction. Importantly, our work shows that the highly coveted  spin-triplet SC order is expected to occur due to the underlying spin-split band
structure and does not require any exotic pairing mechanism; the
conventional retarded phonon-induced attraction is perfectly adequate
for the task. In addition, it is known that the presence of magnetism in a material is not necessarily inimical to superconductivity: high-$T_c$ cuprates and some iron-based superconductors famously emerge from doped antiferromegnets \cite{Dai2015}. Likewise, several ferromagnetic superconductors have been reported in the literature, including a number of $f$-electron compounds such as  CeCu$_2$Si$_2$ \cite{Pfleiderer2009}, as well as UGe$_2$, URhGe, and UCoGe \cite{Aoki2019}.   

It should be noted that Cooper pairing at nonzero center-of-mass
momentum \cite{Chakraborty2024} has been discussed as a possible alternative to the
zero-momentum spin-triplet $p$-wave SC order in
altermagnets. A discovery of such a pair density or
Fulde-Ferrell-Larkin-Ovchinikov (FFLO) state would be perhaps as  exciting as the spin-triplet $p$-wave order, although its utility in 
spintronics or other practical applications is yet to be
understood. In either case search for superconducting altermagnets
holds  promise of exciting future discoveries.

The possibility of induced SC order, of the type useful for spintronic applications,
in a known altermagnet by the
proximity effect with an established superconductor is an interesting
open question. On symmetry grounds one could imagine that proximity-inducing
spin-triplet $p$-wave order using a conventional spin-singlet
superconductor might be challenging. However, a
clever scheme that would leverage interfacial SOC or surface-state
engineering could perhaps bring about such a feat, thus providing a shortcut
towards useful artificially engineered SC altermagnets.

\section*{Acknowledgments}

We are grateful to Tomas Jungwirth,  Vic Law,  Alan MacDonald, Joel
Moore, Andreas Schnyder and  Libor Smejkal for
illuminating discussions and correspondence. This research was
supported in part by NSERC, CIFAR, and the Canada First 
Research Excellence Fund, Quantum Materials and Future Technologies
Program.

\appendix	

\section{Superfluid density tensor}\label{app1}

Various equivalent expressions can be found in the literature for the
superfluid density of a BCS superconductor. Here we use the expression
derived in two different ways in Refs.\ \cite{Sheehy2004,Tummuru2022}, valid for a 2D lattice
superconductor with a normal state dispersion $\xi_\bk$ and gap
function $\Delta_\bk$. Adapted to our model of a $d$-wave altermagnet it
reads
\begin{eqnarray}
[\hat\rho_\sigma]_{\alpha\beta}&=&{e\over 2\hbar^2 S}\sum_\bk \biggl[
  {\partial^2\xi_{\bk\sigma}\over \partial k_\alpha\partial
  k_\beta}\left(1-{\xi_{\bk\sigma}\over E_{\bk\sigma}}\tanh{{1\over
                              2}\beta E_{\bk\sigma}}\right) \nonumber \\
&&-{1\over 2}\beta{\partial \xi_{\bk\sigma}\over\partial k_\alpha}
    {\partial \xi_{\bk\sigma}\over\partial k_\beta}{\rm sech}^2{ {1\over
    2}\beta E_{\bk\sigma}} \biggr]. \label{a1}
\end{eqnarray}
Here the sum is over the Brillouin zone, 
$E_{\bk\sigma}=\sqrt{\xi_{\bk\sigma}^2+\Delta_\sigma^2|S_{\bk\sigma}|^2}$
is the quasiparticle excitation energy and $S$ denotes the area of the
system.

For a fully gapped superconductor the expression simplifies in the
$T\to 0$ limit and gives
\begin{equation}\label{a2}
[\hat\rho_\sigma]_{\alpha\beta}={e\over 2\hbar^2 S}\sum_\bk
  {\partial^2\xi_{\bk\sigma}\over \partial k_\alpha\partial
    k_\beta}\left(1-{\xi_{\bk\sigma}\over E_{\bk\sigma}}\right).
\end{equation}
If we further assume that the gap is small compared to the Fermi
energy and consider electron filling near the bottom of the band, Eq.\
\eqref{a2} can be approximated as   
\begin{equation}\label{a3}
[\hat\rho_\sigma]_{\alpha\beta}\simeq [M_{\sigma}^{-1}]_{\alpha\beta}{e\over 2S}\sum_\bk
\left(1-{\xi_{\bk\sigma}\over |\xi_{\bk\sigma}|}\right).
\end{equation}
Here $[M_\sigma^{-1}]_{\alpha\beta}=\hbar^{-2}
{\partial^2\xi_{\bk\sigma}/\partial k_\alpha\partial k_\beta}$ is the inverse
effective mass tensor and the sum represents the electron density
$n_e$ per spin.

For the altermagnetic energy dispersion that follows from Eq.\
\eqref{h1} we have
\begin{equation}\label{a4}
  M_{\sigma}^{-1}\simeq {2\over \hbar^2}
\begin{pmatrix}
    t+\sigma\eta & 0 \\
    0 & t-\sigma\eta
\end{pmatrix},    
\end{equation}
which then leads directly to Eq.\ \eqref{h17}.

\section{Physics of the $\psi^*_\uparrow\psi_\downarrow$ term}\label{app2}

The $\psi^*_\uparrow\psi_\downarrow$ term in GL free energy Eq.\ \eqref{h20} is symmetry allowed
for the two helical phases listed in Table 1 but is disallowed for the
chiral phases. To see this we consider its behavior under $C_{z4}$
rotation. It is useful to note that the $z$-component of
the total angular momentum of the equal-spin triplet order parameters
is $J_z=(0,0,2,-2)$ for $(p_-^\uparrow,p_+^\downarrow,
p_+^\uparrow,p_-^\downarrow)$, respectively. Under $C_{z4}$ such
order parameters transform as $\psi\to e^{iJ_z\pi/2}\psi$.

It follows that the bilinear term $D(\psi^*_\uparrow\psi_\downarrow +{\rm
  c.c.})$ in  Eq.\ \eqref{h20}  is invariant under $C_{z4}$  if the two order parameters
represent $p_-^\uparrow\otimes p_+^\downarrow$ or
$p_+^\uparrow\otimes p_-^\downarrow$ configurations, that is, the two
helical phases. On the other
hand, the bilinear term is odd under $C_{z4}$ for the two chiral
combinations $p_-^\uparrow\otimes p_-^\downarrow$ and
$p_+^\uparrow\otimes p_+^\downarrow$. In this case the lowest-order
coupling involves a quartic term  $D_4(\psi^{2*}_\uparrow\psi^2_\downarrow +{\rm
  c.c.})$.

In the presence of weak SOC we expect $D\propto \lambda^2$ and $D_4\propto
\lambda^4$ because the two terms represent tunneling of one and two
Cooper pairs between two spin sectors, respectively. In the following
we use the microscopic BdG model defined in Sec.\ II to explicitly
evaluate the Josephson coupling $J=D\psi_0^2$ for the helical
$p_-^\uparrow\otimes p_+^\downarrow$ phase of 
a $d$-wave altermagnet. Specifically, we consider the normal-state lattice
Hamiltonian defined by Eqs.\ \eqref{h1} and \eqref{h2} in the presence
of attractive interaction \eqref{h6}, decoupled in the triplet channel
as shown in Eq.\ \eqref{h9b}.

We obtain $J$ by expanding the free energy of the BdG model
\begin{equation}\label{b1}
F(\varphi)=E_0-{2\over \beta}\sum_{\bk,a}\ln\left[2\cosh{\left({1\over 2}\beta
E_{\bk a}(\varphi)\right)}\right],  
\end{equation}
to leading order in the relative phase $\varphi$ between the two order
parameters. Here $\beta$ denotes the inverse temperature, $E_0$ contains terms
independent of $\varphi$ and $E_{\bk a}(\varphi)$ are two positive eigenvalues of
the $4\times 4$ BdG matrix Hamiltonian Eq.\ \eqref{h11}.  For  the
helical $p_-^\uparrow\otimes p_+^\downarrow$ phase these can be expressed
as 
\begin{equation}\label{b2}
E_{\bk a}^2 (\varphi)=\xi_\bk^2+\eta_\bk^2
+|\lambda_\bk|^2+|\Delta_\bk|^2+2 a D_\bk(\varphi),
\end{equation}
with $a=\pm$,
\begin{equation}\label{b3}
D_{\bk}^2 (\varphi)=\xi_\bk^2(\eta_\bk^2
+|\lambda_\bk|^2)+\Im(\lambda_\bk\Delta_\bk)^2
\end{equation}
and
\begin{eqnarray}\label{b4}
  \xi_\bk&=&-2t(\cos{k_x}+\cos{k_y})-\mu, \nonumber \\
  \eta_\bk&=&-2\eta(\cos{k_x}-\cos{k_y}), \nonumber\\
  \lambda_\bk&=&\lambda_R e^{i\varphi/2}(\sin{k_y}+i\sin{k_x}) \nonumber \\
  \Delta_\bk&=&\Delta_0 (\sin{k_x}+i\sin{k_y}).
\end{eqnarray}

In order to make headway we now focus on the chemical potential near the
bottom of the band and expand $\xi_\bk$ and $\eta_\bk$  in Eq.\ \eqref{b4} to
leading order in small momentum $k$, 
\begin{equation}\label{b5}
\xi_{\bk}\simeq t(k^2-k_F^2), \ \ \ \eta_\bk\simeq \eta(k_x^2-k_y^2),
\end{equation}
where $k_F=\sqrt{(\mu+4t)/t}$ is the Fermi momentum.  The remaining quantities in Eq.\ \eqref{b4}
are similarly expanded near the Fermi level, e.g.\
\begin{equation}\label{b5b}
\Delta_{\bk}\simeq \Delta_0(k_x+ik_y)\approx \Delta_0k_Fe^{i\alpha_\bk},
\end{equation}
where $\alpha_\bk$ is the angle between vector $\bk$ and the $k_x$ axis. 

We note that  the entire $\varphi$ dependence of the free energy is
contained in the last term of Eq.\ \eqref{b3} which is simplified in
this approximation as
\begin{equation}\label{b6}
\Im(\lambda_\bk\Delta_\bk)^2 \simeq
\lambda_R^2\Delta_0^2k_F^4\cos^2(\varphi/2), 
\end{equation}
We expect the leading behavior of the free energy to be of the form
$F(\varphi)\propto \lambda_R^2\cos{\varphi}$. To isolate this term we
now focus on the free energy derivative $F'(\varphi)$ which can be
expressed as
\begin{equation}\label{b7}
F'=-\sum_{\bk,a}{1\over 2E_{\bk a}}{\partial E_{\bk a}^2\over
  \partial\varphi}\tanh{{1\over2}\beta E_{\bk a}}.
\end{equation}
The phase derivative can be written as
\begin{equation}\label{b8}
{\partial E_{\bk a}^2\over  \partial\varphi} =-{1\over
  2}a \lambda_R^2\Delta_0^2k_F^4{\sin\varphi\over D_\bk(\varphi)}.
\end{equation}
Given that this term is proportional to $\lambda_R^2$ and we are
interested in the result to this order we can evaluate the sum in Eq.\
\eqref{b7} by taking $\lambda_R\to 0$ in the rest of the
expression. This simplifies things considerably and we obtain, in the
limit of zero temperature $T$,
\begin{equation}\label{b9}
F'\simeq -{1\over 2} \lambda_R^2\Delta_0^2k_F^4
\sum_{\bk,a}{a \sin{\varphi}\over
  |\xi_\bk\eta_\bk|\sqrt{(\xi_\bk+a \eta_\bk)^2+\Delta_0^2}}.
\end{equation}
Hence we see that to leading order $F'\propto \lambda_R^2\sin{\varphi}$, as
expected.

To find the proportionality constant we must evaluate the remaining
sum over $\bk$. The leading contribution can be estimated already in
the limit of zero spin splitting $\eta$. Taking the limit $\eta\to 0$
we find
\begin{equation}\label{b10}
F'\simeq \lambda_R^2\Delta_0^2k_F^4 \sin{\varphi}
\sum_{\bk}{1\over
  (\xi_\bk^2+\Delta_0^2)^{3/2}}.
\end{equation}
We evaluate the sum and arrive at
\begin{equation}\label{b11}
F'\simeq {\lambda_R^2k_F^4\over 4\pi t} \sin{\varphi}.
\end{equation}
The expression is valid for $\lambda_R\lesssim \Delta_0$ and for the
chemical potential near the bottom of the band such that the expansion of $\xi_\bk$ in Eq.\
\eqref{b5} is accurate. The last condition is met when $k_F\lesssim 1$.

We thus conclude that to the leading order in the phase difference the free energy can be
written as  
\begin{equation}\label{b12}
F(\varphi)\simeq F_0 -J \cos{\varphi}, \ \ \ J \simeq
{\lambda_R^2k_F^4\over 4\pi t}, 
\end{equation}
where $F_0$ is independent of $\varphi$. We note that because of the
$4\pi$ factor in the denominator even a fairly significant
SOC strength $\lambda_R\sim t$ will tend to yield relatively small
values of $J$. In addition, $J$ can be further suppressed by going to
low electron density limit where $k_F\ll 1$. We conclude that
generically one may expect the linear coupling between spin-up and
spin-down order parameters to be parametrically small
compared to other energy scales in the problem.

We note that a similar
calculation can be performed for the
$p_+^\uparrow\otimes p_-^\downarrow$ helical phase. In this case
$\Im(\lambda_\bk\Delta_\bk)$ in Eq.\ \eqref{b3} is replaced by
$\Im(\lambda_\bk\Delta_\bk^*)$.  One obtains the
same expression \eqref{b12} for the free energy except with a
modified $J\simeq -\lambda_R^2k_F^6/192t$.  
Finally, the  calculation can be performed for the two chiral phases
listed in Table 1. In
this case the form of the BdG energy eigenvalues is different from Eq.\
\eqref{b2} and the calculation to the same order gives $J=0$,
consistent with the expectation that the leading contribution is
fourth order in $\lambda_R$. 

\section{Spin current with random $J$}
\begin{figure}[t]
  \includegraphics[width=8.5cm]{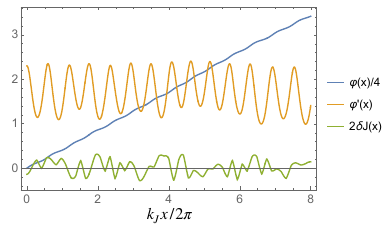}
  \caption{Spin current in a long wire calculated from the GL theory with a spatially varying random internal Josephson coupling described in the text. We used $M=42$ random values with $\langle\delta J^2\rangle^{1/2}=0.3J_0$ and $\varphi'(0)=2.3$.}
  \label{fig7}
\end{figure}

To model disorder we consider free energy density Eq.\ \eqref{h22} with the internal Josephson coupling containing a spatially varying component  
\begin{equation}\label{c1}
J(x)=J_0+\delta J(x).
\end{equation}
Here $\delta J(x)$ is generated by smoothly interpolating $M$ random numbers $w(x_j)$ drawn from a uniform distribution $[-r,r]$ at spatial positions $x_j=jL/M$ with $j=1,\dots M$. The corresponding Euler-Lagrange equation  
\begin{equation}\label{b2}
{d^2\varphi\over d x^2} +k_J^2[1+\zeta(x)]\sin{\varphi}=0, 
\end{equation}
with $\zeta(x)=\delta J(x)/J_0$ is then solved numerically subject to boundary conditions at $x=0$ just like in Sec.\ IV.B. A representative solution is displayed in Fig.\ \ref{fig7}. We find that non-decaying oscillatory solutions of the type displayed generally persist for small-amplitude disorder. When the injected current density is close to the critical value $j_c^s$ we sometimes find that a large fluctuation in $\delta J$ can switch the behavior between two types displayed in Figs.\ \ref{fig4}(a) and (c). Nevertheless, the current remains undamped in all cases.

\bibliography{spin}

\begin{thebibliography}{70}%
\makeatletter
\providecommand \@ifxundefined [1]{%
 \@ifx{#1\undefined}
}%
\providecommand \@ifnum [1]{%
 \ifnum #1\expandafter \@firstoftwo
 \else \expandafter \@secondoftwo
 \fi
}%
\providecommand \@ifx [1]{%
 \ifx #1\expandafter \@firstoftwo
 \else \expandafter \@secondoftwo
 \fi
}%
\providecommand \natexlab [1]{#1}%
\providecommand \enquote  [1]{``#1''}%
\providecommand \bibnamefont  [1]{#1}%
\providecommand \bibfnamefont [1]{#1}%
\providecommand \citenamefont [1]{#1}%
\providecommand \href@noop [0]{\@secondoftwo}%
\providecommand \href [0]{\begingroup \@sanitize@url \@href}%
\providecommand \@href[1]{\@@startlink{#1}\@@href}%
\providecommand \@@href[1]{\endgroup#1\@@endlink}%
\providecommand \@sanitize@url [0]{\catcode `\\12\catcode `\$12\catcode
  `\&12\catcode `\#12\catcode `\^12\catcode `\_12\catcode `\%12\relax}%
\providecommand \@@startlink[1]{}%
\providecommand \@@endlink[0]{}%
\providecommand \url  [0]{\begingroup\@sanitize@url \@url }%
\providecommand \@url [1]{\endgroup\@href {#1}{\urlprefix }}%
\providecommand \urlprefix  [0]{URL }%
\providecommand \Eprint [0]{\href }%
\providecommand \doibase [0]{http://dx.doi.org/}%
\providecommand \selectlanguage [0]{\@gobble}%
\providecommand \bibinfo  [0]{\@secondoftwo}%
\providecommand \bibfield  [0]{\@secondoftwo}%
\providecommand \translation [1]{[#1]}%
\providecommand \BibitemOpen [0]{}%
\providecommand \bibitemStop [0]{}%
\providecommand \bibitemNoStop [0]{.\EOS\space}%
\providecommand \EOS [0]{\spacefactor3000\relax}%
\providecommand \BibitemShut  [1]{\csname bibitem#1\endcsname}%
\let\auto@bib@innerbib\@empty
\bibitem [{\citenamefont {Broom}(1961)}]{Broom1961}%
  \BibitemOpen
  \bibfield  {author} {\bibinfo {author} {\bibfnamefont {R.~F.}\ \bibnamefont
  {Broom}},\ }\bibfield  {title} {\enquote {\bibinfo {title} {An upper limit
  for the resistivity of a superconducting film},}\ }\href {\doibase
  10.1038/190992b0} {\bibfield  {journal} {\bibinfo  {journal} {Nature}\
  }\textbf {\bibinfo {volume} {190}},\ \bibinfo {pages} {992--993} (\bibinfo
  {year} {1961})}\BibitemShut {NoStop}%
\bibitem [{\citenamefont {File}\ and\ \citenamefont {Mills}(1963)}]{File1963}%
  \BibitemOpen
  \bibfield  {author} {\bibinfo {author} {\bibfnamefont {J.}~\bibnamefont
  {File}}\ and\ \bibinfo {author} {\bibfnamefont {R.~G.}\ \bibnamefont
  {Mills}},\ }\bibfield  {title} {\enquote {\bibinfo {title} {Observation of
  persistent current in a superconducting solenoid},}\ }\href {\doibase
  10.1103/PhysRevLett.10.93} {\bibfield  {journal} {\bibinfo  {journal} {Phys.
  Rev. Lett.}\ }\textbf {\bibinfo {volume} {10}},\ \bibinfo {pages} {93--96}
  (\bibinfo {year} {1963})}\BibitemShut {NoStop}%
\bibitem [{\citenamefont {\ifmmode \check{Z}\else
  \v{Z}\fi{}uti\ifmmode~\acute{c}\else \'{c}\fi{}}\ \emph
  {et~al.}(2004)\citenamefont {\ifmmode \check{Z}\else
  \v{Z}\fi{}uti\ifmmode~\acute{c}\else \'{c}\fi{}}, \citenamefont {Fabian},\
  and\ \citenamefont {Das~Sarma}}]{Fabian2004}%
  \BibitemOpen
  \bibfield  {author} {\bibinfo {author} {\bibfnamefont {Igor}\ \bibnamefont
  {\ifmmode \check{Z}\else \v{Z}\fi{}uti\ifmmode~\acute{c}\else \'{c}\fi{}}},
  \bibinfo {author} {\bibfnamefont {Jaroslav}\ \bibnamefont {Fabian}}, \ and\
  \bibinfo {author} {\bibfnamefont {S.}~\bibnamefont {Das~Sarma}},\ }\bibfield
  {title} {\enquote {\bibinfo {title} {Spintronics: Fundamentals and
  applications},}\ }\href {\doibase 10.1103/RevModPhys.76.323} {\bibfield
  {journal} {\bibinfo  {journal} {Rev. Mod. Phys.}\ }\textbf {\bibinfo {volume}
  {76}},\ \bibinfo {pages} {323--410} (\bibinfo {year} {2004})}\BibitemShut
  {NoStop}%
\bibitem [{\citenamefont {Sinova}\ \emph {et~al.}(2015)\citenamefont {Sinova},
  \citenamefont {Valenzuela}, \citenamefont {Wunderlich}, \citenamefont
  {Back},\ and\ \citenamefont {Jungwirth}}]{Sinova2015}%
  \BibitemOpen
  \bibfield  {author} {\bibinfo {author} {\bibfnamefont {Jairo}\ \bibnamefont
  {Sinova}}, \bibinfo {author} {\bibfnamefont {Sergio~O.}\ \bibnamefont
  {Valenzuela}}, \bibinfo {author} {\bibfnamefont {J.}~\bibnamefont
  {Wunderlich}}, \bibinfo {author} {\bibfnamefont {C.~H.}\ \bibnamefont
  {Back}}, \ and\ \bibinfo {author} {\bibfnamefont {T.}~\bibnamefont
  {Jungwirth}},\ }\bibfield  {title} {\enquote {\bibinfo {title} {Spin hall
  effects},}\ }\href {\doibase 10.1103/RevModPhys.87.1213} {\bibfield
  {journal} {\bibinfo  {journal} {Rev. Mod. Phys.}\ }\textbf {\bibinfo {volume}
  {87}},\ \bibinfo {pages} {1213--1260} (\bibinfo {year} {2015})}\BibitemShut
  {NoStop}%
\bibitem [{\citenamefont {Linder}\ and\ \citenamefont
  {Robinson}(2015)}]{Robinson2015}%
  \BibitemOpen
  \bibfield  {author} {\bibinfo {author} {\bibfnamefont {Jacob}\ \bibnamefont
  {Linder}}\ and\ \bibinfo {author} {\bibfnamefont {Jason W.~A.}\ \bibnamefont
  {Robinson}},\ }\bibfield  {title} {\enquote {\bibinfo {title}
  {Superconducting spintronics},}\ }\href {\doibase 10.1038/nphys3242}
  {\bibfield  {journal} {\bibinfo  {journal} {Nature Physics}\ }\textbf
  {\bibinfo {volume} {11}},\ \bibinfo {pages} {307--315} (\bibinfo {year}
  {2015})}\BibitemShut {NoStop}%
\bibitem [{\citenamefont {Eschrig}(2015)}]{Eschrig2015}%
  \BibitemOpen
  \bibfield  {author} {\bibinfo {author} {\bibfnamefont {Matthias}\
  \bibnamefont {Eschrig}},\ }\bibfield  {title} {\enquote {\bibinfo {title}
  {Spin-polarized supercurrents for spintronics: a review of current
  progress},}\ }\href {\doibase 10.1088/0034-4885/78/10/104501} {\bibfield
  {journal} {\bibinfo  {journal} {Reports on Progress in Physics}\ }\textbf
  {\bibinfo {volume} {78}},\ \bibinfo {pages} {104501} (\bibinfo {year}
  {2015})}\BibitemShut {NoStop}%
\bibitem [{\citenamefont {Jeon}\ \emph {et~al.}(2018)\citenamefont {Jeon},
  \citenamefont {Ciccarelli}, \citenamefont {Ferguson}, \citenamefont
  {Kurebayashi}, \citenamefont {Cohen}, \citenamefont {Montiel}, \citenamefont
  {Eschrig}, \citenamefont {Robinson},\ and\ \citenamefont
  {Blamire}}]{Blamire2018}%
  \BibitemOpen
  \bibfield  {author} {\bibinfo {author} {\bibfnamefont {Kun-Rok}\ \bibnamefont
  {Jeon}}, \bibinfo {author} {\bibfnamefont {Chiara}\ \bibnamefont
  {Ciccarelli}}, \bibinfo {author} {\bibfnamefont {Andrew~J.}\ \bibnamefont
  {Ferguson}}, \bibinfo {author} {\bibfnamefont {Hidekazu}\ \bibnamefont
  {Kurebayashi}}, \bibinfo {author} {\bibfnamefont {Lesley~F.}\ \bibnamefont
  {Cohen}}, \bibinfo {author} {\bibfnamefont {Xavier}\ \bibnamefont {Montiel}},
  \bibinfo {author} {\bibfnamefont {Matthias}\ \bibnamefont {Eschrig}},
  \bibinfo {author} {\bibfnamefont {Jason W.~A.}\ \bibnamefont {Robinson}}, \
  and\ \bibinfo {author} {\bibfnamefont {Mark~G.}\ \bibnamefont {Blamire}},\
  }\bibfield  {title} {\enquote {\bibinfo {title} {Enhanced spin pumping into
  superconductors provides evidence for superconducting pure spin currents},}\
  }\href {\doibase 10.1038/s41563-018-0058-9} {\bibfield  {journal} {\bibinfo
  {journal} {Nature Materials}\ }\textbf {\bibinfo {volume} {17}},\ \bibinfo
  {pages} {499--503} (\bibinfo {year} {2018})}\BibitemShut {NoStop}%
\bibitem [{\citenamefont {Satchell}\ and\ \citenamefont
  {Birge}(2018)}]{Norman2018}%
  \BibitemOpen
  \bibfield  {author} {\bibinfo {author} {\bibfnamefont {Nathan}\ \bibnamefont
  {Satchell}}\ and\ \bibinfo {author} {\bibfnamefont {Norman~O.}\ \bibnamefont
  {Birge}},\ }\bibfield  {title} {\enquote {\bibinfo {title} {Supercurrent in
  ferromagnetic josephson junctions with heavy metal interlayers},}\ }\href
  {\doibase 10.1103/PhysRevB.97.214509} {\bibfield  {journal} {\bibinfo
  {journal} {Phys. Rev. B}\ }\textbf {\bibinfo {volume} {97}},\ \bibinfo
  {pages} {214509} (\bibinfo {year} {2018})}\BibitemShut {NoStop}%
\bibitem [{\citenamefont {Jeon}\ \emph {et~al.}(2019)\citenamefont {Jeon},
  \citenamefont {Ciccarelli}, \citenamefont {Kurebayashi}, \citenamefont
  {Cohen}, \citenamefont {Montiel}, \citenamefont {Eschrig}, \citenamefont
  {Komori}, \citenamefont {Robinson},\ and\ \citenamefont
  {Blamire}}]{Blamire2019}%
  \BibitemOpen
  \bibfield  {author} {\bibinfo {author} {\bibfnamefont {Kun-Rok}\ \bibnamefont
  {Jeon}}, \bibinfo {author} {\bibfnamefont {Chiara}\ \bibnamefont
  {Ciccarelli}}, \bibinfo {author} {\bibfnamefont {Hidekazu}\ \bibnamefont
  {Kurebayashi}}, \bibinfo {author} {\bibfnamefont {Lesley~F.}\ \bibnamefont
  {Cohen}}, \bibinfo {author} {\bibfnamefont {Xavier}\ \bibnamefont {Montiel}},
  \bibinfo {author} {\bibfnamefont {Matthias}\ \bibnamefont {Eschrig}},
  \bibinfo {author} {\bibfnamefont {Sachio}\ \bibnamefont {Komori}}, \bibinfo
  {author} {\bibfnamefont {Jason W.~A.}\ \bibnamefont {Robinson}}, \ and\
  \bibinfo {author} {\bibfnamefont {Mark~G.}\ \bibnamefont {Blamire}},\
  }\bibfield  {title} {\enquote {\bibinfo {title} {Exchange-field enhancement
  of superconducting spin pumping},}\ }\href {\doibase
  10.1103/PhysRevB.99.024507} {\bibfield  {journal} {\bibinfo  {journal} {Phys.
  Rev. B}\ }\textbf {\bibinfo {volume} {99}},\ \bibinfo {pages} {024507}
  (\bibinfo {year} {2019})}\BibitemShut {NoStop}%
\bibitem [{\citenamefont {Mart\'{\i}nez}\ \emph {et~al.}(2020)\citenamefont
  {Mart\'{\i}nez}, \citenamefont {H\"ogl}, \citenamefont {Gonz\'alez-Ruano},
  \citenamefont {Cascales}, \citenamefont {Tiusan}, \citenamefont {Lu},
  \citenamefont {Hehn}, \citenamefont {Matos-Abiague}, \citenamefont {Fabian},
  \citenamefont {\ifmmode \check{Z}\else \v{Z}\fi{}uti\ifmmode~\acute{c}\else
  \'{c}\fi{}},\ and\ \citenamefont {Aliev}}]{Farkhad2020}%
  \BibitemOpen
  \bibfield  {author} {\bibinfo {author} {\bibfnamefont {Isidoro}\ \bibnamefont
  {Mart\'{\i}nez}}, \bibinfo {author} {\bibfnamefont {Petra}\ \bibnamefont
  {H\"ogl}}, \bibinfo {author} {\bibfnamefont {C\'esar}\ \bibnamefont
  {Gonz\'alez-Ruano}}, \bibinfo {author} {\bibfnamefont {Juan~Pedro}\
  \bibnamefont {Cascales}}, \bibinfo {author} {\bibfnamefont {Coriolan}\
  \bibnamefont {Tiusan}}, \bibinfo {author} {\bibfnamefont {Yuan}\ \bibnamefont
  {Lu}}, \bibinfo {author} {\bibfnamefont {Michel}\ \bibnamefont {Hehn}},
  \bibinfo {author} {\bibfnamefont {Alex}\ \bibnamefont {Matos-Abiague}},
  \bibinfo {author} {\bibfnamefont {Jaroslav}\ \bibnamefont {Fabian}}, \bibinfo
  {author} {\bibfnamefont {Igor}\ \bibnamefont {\ifmmode \check{Z}\else
  \v{Z}\fi{}uti\ifmmode~\acute{c}\else \'{c}\fi{}}}, \ and\ \bibinfo {author}
  {\bibfnamefont {Farkhad~G.}\ \bibnamefont {Aliev}},\ }\bibfield  {title}
  {\enquote {\bibinfo {title} {Interfacial spin-orbit coupling: A platform for
  superconducting spintronics},}\ }\href {\doibase
  10.1103/PhysRevApplied.13.014030} {\bibfield  {journal} {\bibinfo  {journal}
  {Phys. Rev. Appl.}\ }\textbf {\bibinfo {volume} {13}},\ \bibinfo {pages}
  {014030} (\bibinfo {year} {2020})}\BibitemShut {NoStop}%
\bibitem [{\citenamefont {Jeon}\ \emph {et~al.}(2020)\citenamefont {Jeon},
  \citenamefont {Montiel}, \citenamefont {Komori}, \citenamefont {Ciccarelli},
  \citenamefont {Haigh}, \citenamefont {Kurebayashi}, \citenamefont {Cohen},
  \citenamefont {Chan}, \citenamefont {Stenning}, \citenamefont {Lee},
  \citenamefont {Eschrig}, \citenamefont {Blamire},\ and\ \citenamefont
  {Robinson}}]{Robinson2020}%
  \BibitemOpen
  \bibfield  {author} {\bibinfo {author} {\bibfnamefont {Kun-Rok}\ \bibnamefont
  {Jeon}}, \bibinfo {author} {\bibfnamefont {Xavier}\ \bibnamefont {Montiel}},
  \bibinfo {author} {\bibfnamefont {Sachio}\ \bibnamefont {Komori}}, \bibinfo
  {author} {\bibfnamefont {Chiara}\ \bibnamefont {Ciccarelli}}, \bibinfo
  {author} {\bibfnamefont {James}\ \bibnamefont {Haigh}}, \bibinfo {author}
  {\bibfnamefont {Hidekazu}\ \bibnamefont {Kurebayashi}}, \bibinfo {author}
  {\bibfnamefont {Lesley~F.}\ \bibnamefont {Cohen}}, \bibinfo {author}
  {\bibfnamefont {Alex~K.}\ \bibnamefont {Chan}}, \bibinfo {author}
  {\bibfnamefont {Kilian~D.}\ \bibnamefont {Stenning}}, \bibinfo {author}
  {\bibfnamefont {Chang-Min}\ \bibnamefont {Lee}}, \bibinfo {author}
  {\bibfnamefont {Matthias}\ \bibnamefont {Eschrig}}, \bibinfo {author}
  {\bibfnamefont {Mark~G.}\ \bibnamefont {Blamire}}, \ and\ \bibinfo {author}
  {\bibfnamefont {Jason W.~A.}\ \bibnamefont {Robinson}},\ }\bibfield  {title}
  {\enquote {\bibinfo {title} {Tunable pure spin supercurrents and the
  demonstration of their gateability in a spin-wave device},}\ }\href {\doibase
  10.1103/PhysRevX.10.031020} {\bibfield  {journal} {\bibinfo  {journal} {Phys.
  Rev. X}\ }\textbf {\bibinfo {volume} {10}},\ \bibinfo {pages} {031020}
  (\bibinfo {year} {2020})}\BibitemShut {NoStop}%
\bibitem [{\citenamefont {He}\ \emph {et~al.}(2019)\citenamefont {He},
  \citenamefont {Hiroki}, \citenamefont {Hamamoto},\ and\ \citenamefont
  {Nagaosa}}]{Nagaosa2019}%
  \BibitemOpen
  \bibfield  {author} {\bibinfo {author} {\bibfnamefont {James~Jun}\
  \bibnamefont {He}}, \bibinfo {author} {\bibfnamefont {Kanta}\ \bibnamefont
  {Hiroki}}, \bibinfo {author} {\bibfnamefont {Keita}\ \bibnamefont
  {Hamamoto}}, \ and\ \bibinfo {author} {\bibfnamefont {Naoto}\ \bibnamefont
  {Nagaosa}},\ }\bibfield  {title} {\enquote {\bibinfo {title} {Spin
  supercurrent in two-dimensional superconductors with rashba spin-orbit
  interaction},}\ }\href {\doibase 10.1038/s42005-019-0230-9} {\bibfield
  {journal} {\bibinfo  {journal} {Communications Physics}\ }\textbf {\bibinfo
  {volume} {2}},\ \bibinfo {pages} {128} (\bibinfo {year} {2019})}\BibitemShut
  {NoStop}%
\bibitem [{\citenamefont {Silaev}\ \emph {et~al.}(2020)\citenamefont {Silaev},
  \citenamefont {Bobkova},\ and\ \citenamefont {Bobkov}}]{Silaev2020}%
  \BibitemOpen
  \bibfield  {author} {\bibinfo {author} {\bibfnamefont {M.~A.}\ \bibnamefont
  {Silaev}}, \bibinfo {author} {\bibfnamefont {I.~V.}\ \bibnamefont {Bobkova}},
  \ and\ \bibinfo {author} {\bibfnamefont {A.~M.}\ \bibnamefont {Bobkov}},\
  }\bibfield  {title} {\enquote {\bibinfo {title} {Odd triplet
  superconductivity induced by a moving condensate},}\ }\href {\doibase
  10.1103/PhysRevB.102.100507} {\bibfield  {journal} {\bibinfo  {journal}
  {Phys. Rev. B}\ }\textbf {\bibinfo {volume} {102}},\ \bibinfo {pages}
  {100507} (\bibinfo {year} {2020})}\BibitemShut {NoStop}%
\bibitem [{\citenamefont {Montiel}\ and\ \citenamefont
  {Eschrig}(2023)}]{Montiel2023}%
  \BibitemOpen
  \bibfield  {author} {\bibinfo {author} {\bibfnamefont {X.}~\bibnamefont
  {Montiel}}\ and\ \bibinfo {author} {\bibfnamefont {M.}~\bibnamefont
  {Eschrig}},\ }\bibfield  {title} {\enquote {\bibinfo {title} {Spin current
  injection via equal-spin cooper pairs in ferromagnet/superconductor
  heterostructures},}\ }\href {\doibase 10.1103/PhysRevB.107.094513} {\bibfield
   {journal} {\bibinfo  {journal} {Phys. Rev. B}\ }\textbf {\bibinfo {volume}
  {107}},\ \bibinfo {pages} {094513} (\bibinfo {year} {2023})}\BibitemShut
  {NoStop}%
\bibitem [{\citenamefont {Chourasia}\ \emph {et~al.}(2023)\citenamefont
  {Chourasia}, \citenamefont {Kamra}, \citenamefont {Bobkova},\ and\
  \citenamefont {Kamra}}]{Kamra2023}%
  \BibitemOpen
  \bibfield  {author} {\bibinfo {author} {\bibfnamefont {Simran}\ \bibnamefont
  {Chourasia}}, \bibinfo {author} {\bibfnamefont {Lina~Johnsen}\ \bibnamefont
  {Kamra}}, \bibinfo {author} {\bibfnamefont {Irina~V.}\ \bibnamefont
  {Bobkova}}, \ and\ \bibinfo {author} {\bibfnamefont {Akashdeep}\ \bibnamefont
  {Kamra}},\ }\bibfield  {title} {\enquote {\bibinfo {title} {Generation of
  spin-triplet cooper pairs via a canted antiferromagnet},}\ }\href {\doibase
  10.1103/PhysRevB.108.064515} {\bibfield  {journal} {\bibinfo  {journal}
  {Phys. Rev. B}\ }\textbf {\bibinfo {volume} {108}},\ \bibinfo {pages}
  {064515} (\bibinfo {year} {2023})}\BibitemShut {NoStop}%
\bibitem [{\citenamefont {Bobkov}\ \emph {et~al.}(2024)\citenamefont {Bobkov},
  \citenamefont {Bobkov},\ and\ \citenamefont {Bobkova}}]{Bobkova2024}%
  \BibitemOpen
  \bibfield  {author} {\bibinfo {author} {\bibfnamefont {G.~A.}\ \bibnamefont
  {Bobkov}}, \bibinfo {author} {\bibfnamefont {A.~M.}\ \bibnamefont {Bobkov}},
  \ and\ \bibinfo {author} {\bibfnamefont {I.~V.}\ \bibnamefont {Bobkova}},\
  }\bibfield  {title} {\enquote {\bibinfo {title} {Spin supercurrent in
  superconductor/ferromagnet van der waals heterostructures},}\ }\href
  {\doibase 10.1103/PhysRevB.110.104506} {\bibfield  {journal} {\bibinfo
  {journal} {Phys. Rev. B}\ }\textbf {\bibinfo {volume} {110}},\ \bibinfo
  {pages} {104506} (\bibinfo {year} {2024})}\BibitemShut {NoStop}%
\bibitem [{\citenamefont {Ahn}\ \emph {et~al.}(2019)\citenamefont {Ahn},
  \citenamefont {Hariki}, \citenamefont {Lee},\ and\ \citenamefont
  {Kune{\v{s}}}}]{ahn2019antiferromagnetism}%
  \BibitemOpen
  \bibfield  {author} {\bibinfo {author} {\bibfnamefont {Kyo-Hoon}\
  \bibnamefont {Ahn}}, \bibinfo {author} {\bibfnamefont {Atsushi}\ \bibnamefont
  {Hariki}}, \bibinfo {author} {\bibfnamefont {Kwan-Woo}\ \bibnamefont {Lee}},
  \ and\ \bibinfo {author} {\bibfnamefont {Jan}\ \bibnamefont {Kune{\v{s}}}},\
  }\bibfield  {title} {\enquote {\bibinfo {title} {Antiferromagnetism in ruo 2
  as d-wave pomeranchuk instability},}\ }\href@noop {} {\bibfield  {journal}
  {\bibinfo  {journal} {Physical Review B}\ }\textbf {\bibinfo {volume} {99}},\
  \bibinfo {pages} {184432} (\bibinfo {year} {2019})}\BibitemShut {NoStop}%
\bibitem [{\citenamefont {Hayami}\ \emph {et~al.}(2019)\citenamefont {Hayami},
  \citenamefont {Yanagi},\ and\ \citenamefont {Kusunose}}]{Hayami2020}%
  \BibitemOpen
  \bibfield  {author} {\bibinfo {author} {\bibfnamefont {Satoru}\ \bibnamefont
  {Hayami}}, \bibinfo {author} {\bibfnamefont {Yuki}\ \bibnamefont {Yanagi}}, \
  and\ \bibinfo {author} {\bibfnamefont {Hiroaki}\ \bibnamefont {Kusunose}},\
  }\bibfield  {title} {\enquote {\bibinfo {title} {Momentum-dependent spin
  splitting by collinear antiferromagnetic ordering},}\ }\href {\doibase
  10.7566/JPSJ.88.123702} {\bibfield  {journal} {\bibinfo  {journal} {Journal
  of the Physical Society of Japan}\ }\textbf {\bibinfo {volume} {88}},\
  \bibinfo {pages} {123702} (\bibinfo {year} {2019})}\BibitemShut {NoStop}%
\bibitem [{\citenamefont {\ifmmode~\check{S}\else \v{S}\fi{}mejkal}\ \emph
  {et~al.}(2022{\natexlab{a}})\citenamefont {\ifmmode~\check{S}\else
  \v{S}\fi{}mejkal}, \citenamefont {Sinova},\ and\ \citenamefont
  {Jungwirth}}]{Smejkal2022a}%
  \BibitemOpen
  \bibfield  {author} {\bibinfo {author} {\bibfnamefont {Libor}\ \bibnamefont
  {\ifmmode~\check{S}\else \v{S}\fi{}mejkal}}, \bibinfo {author} {\bibfnamefont
  {Jairo}\ \bibnamefont {Sinova}}, \ and\ \bibinfo {author} {\bibfnamefont
  {Tomas}\ \bibnamefont {Jungwirth}},\ }\bibfield  {title} {\enquote {\bibinfo
  {title} {Beyond conventional ferromagnetism and antiferromagnetism: A phase
  with nonrelativistic spin and crystal rotation symmetry},}\ }\href {\doibase
  10.1103/PhysRevX.12.031042} {\bibfield  {journal} {\bibinfo  {journal} {Phys.
  Rev. X}\ }\textbf {\bibinfo {volume} {12}},\ \bibinfo {pages} {031042}
  (\bibinfo {year} {2022}{\natexlab{a}})}\BibitemShut {NoStop}%
\bibitem [{\citenamefont {\ifmmode~\check{S}\else \v{S}\fi{}mejkal}\ \emph
  {et~al.}(2022{\natexlab{b}})\citenamefont {\ifmmode~\check{S}\else
  \v{S}\fi{}mejkal}, \citenamefont {Sinova},\ and\ \citenamefont
  {Jungwirth}}]{Smejkal2022b}%
  \BibitemOpen
  \bibfield  {author} {\bibinfo {author} {\bibfnamefont {Libor}\ \bibnamefont
  {\ifmmode~\check{S}\else \v{S}\fi{}mejkal}}, \bibinfo {author} {\bibfnamefont
  {Jairo}\ \bibnamefont {Sinova}}, \ and\ \bibinfo {author} {\bibfnamefont
  {Tomas}\ \bibnamefont {Jungwirth}},\ }\bibfield  {title} {\enquote {\bibinfo
  {title} {Emerging research landscape of altermagnetism},}\ }\href {\doibase
  10.1103/PhysRevX.12.040501} {\bibfield  {journal} {\bibinfo  {journal} {Phys.
  Rev. X}\ }\textbf {\bibinfo {volume} {12}},\ \bibinfo {pages} {040501}
  (\bibinfo {year} {2022}{\natexlab{b}})}\BibitemShut {NoStop}%
\bibitem [{\citenamefont {Mazin}(2022)}]{Mazin2022}%
  \BibitemOpen
  \bibfield  {author} {\bibinfo {author} {\bibfnamefont {Igor}\ \bibnamefont
  {Mazin}} (\bibinfo {collaboration} {The PRX Editors}),\ }\bibfield  {title}
  {\enquote {\bibinfo {title} {Editorial: Altermagnetism---a new punch line of
  fundamental magnetism},}\ }\href {\doibase 10.1103/PhysRevX.12.040002}
  {\bibfield  {journal} {\bibinfo  {journal} {Phys. Rev. X}\ }\textbf {\bibinfo
  {volume} {12}},\ \bibinfo {pages} {040002} (\bibinfo {year}
  {2022})}\BibitemShut {NoStop}%
\bibitem [{\citenamefont {Jiang}\ \emph {et~al.}(2025)\citenamefont {Jiang},
  \citenamefont {Hu}, \citenamefont {Bai}, \citenamefont {Song}, \citenamefont
  {Mu}, \citenamefont {Qu}, \citenamefont {Li}, \citenamefont {Zhu},
  \citenamefont {Pi}, \citenamefont {Wei} \emph {et~al.}}]{Jiang2024}%
  \BibitemOpen
  \bibfield  {author} {\bibinfo {author} {\bibfnamefont {Bei}\ \bibnamefont
  {Jiang}}, \bibinfo {author} {\bibfnamefont {Mingzhe}\ \bibnamefont {Hu}},
  \bibinfo {author} {\bibfnamefont {Jianli}\ \bibnamefont {Bai}}, \bibinfo
  {author} {\bibfnamefont {Ziyin}\ \bibnamefont {Song}}, \bibinfo {author}
  {\bibfnamefont {Chao}\ \bibnamefont {Mu}}, \bibinfo {author} {\bibfnamefont
  {Gexing}\ \bibnamefont {Qu}}, \bibinfo {author} {\bibfnamefont {Wan}\
  \bibnamefont {Li}}, \bibinfo {author} {\bibfnamefont {Wenliang}\ \bibnamefont
  {Zhu}}, \bibinfo {author} {\bibfnamefont {Hanqi}\ \bibnamefont {Pi}},
  \bibinfo {author} {\bibfnamefont {Zhongxu}\ \bibnamefont {Wei}},  \emph
  {et~al.},\ }\bibfield  {title} {\enquote {\bibinfo {title} {A metallic
  room-temperature d-wave altermagnet},}\ }\href@noop {} {\bibfield  {journal}
  {\bibinfo  {journal} {Nature Physics}\ ,\ \bibinfo {pages} {1--6}} (\bibinfo
  {year} {2025})}\BibitemShut {NoStop}%
\bibitem [{\citenamefont {Kitz}(1965)}]{ssgs1}%
  \BibitemOpen
  \bibfield  {author} {\bibinfo {author} {\bibfnamefont {A.}~\bibnamefont
  {Kitz}},\ }\bibfield  {title} {\enquote {\bibinfo {title} {Über die
  symmetriegruppen von spinverteilungen von},}\ }\href {\doibase
  https://doi.org/10.1002/pssb.2220100206} {\bibfield  {journal} {\bibinfo
  {journal} {physica status solidi (b)}\ }\textbf {\bibinfo {volume} {10}},\
  \bibinfo {pages} {455--466} (\bibinfo {year} {1965})}\BibitemShut {NoStop}%
\bibitem [{\citenamefont {Brinkman}\ and\ \citenamefont
  {Elliott}(1966{\natexlab{a}})}]{ssgs2}%
  \BibitemOpen
  \bibfield  {author} {\bibinfo {author} {\bibfnamefont {WF}~\bibnamefont
  {Brinkman}}\ and\ \bibinfo {author} {\bibfnamefont {Roger~James}\
  \bibnamefont {Elliott}},\ }\bibfield  {title} {\enquote {\bibinfo {title}
  {Theory of spin-space groups},}\ }\href
  {https://royalsocietypublishing.org/doi/abs/10.1098/rspa.1966.0211?casa_token=_dguHmeOPwcAAAAA:0x1b_D5dWdPXNszJYu5xORExOCfdglsGcFV5_4A-Mo0C6g2rrRbZE4dUGeHwrAeJOaE-sgvD8tvs6p8F}
  {\bibfield  {journal} {\bibinfo  {journal} {Proceedings of the Royal Society
  of London. Series A. Mathematical and Physical Sciences}\ }\textbf {\bibinfo
  {volume} {294}},\ \bibinfo {pages} {343--358} (\bibinfo {year}
  {1966}{\natexlab{a}})}\BibitemShut {NoStop}%
\bibitem [{\citenamefont {Brinkman}\ and\ \citenamefont
  {Elliott}(1966{\natexlab{b}})}]{ssgs3}%
  \BibitemOpen
  \bibfield  {author} {\bibinfo {author} {\bibfnamefont {W.}~\bibnamefont
  {Brinkman}}\ and\ \bibinfo {author} {\bibfnamefont {R.~J.}\ \bibnamefont
  {Elliott}},\ }\bibfield  {title} {\enquote {\bibinfo {title} {{Space Group
  Theory for Spin Waves}},}\ }\href {\doibase 10.1063/1.1708514} {\bibfield
  {journal} {\bibinfo  {journal} {Journal of Applied Physics}\ }\textbf
  {\bibinfo {volume} {37}},\ \bibinfo {pages} {1457--1459} (\bibinfo {year}
  {1966}{\natexlab{b}})}\BibitemShut {NoStop}%
\bibitem [{\citenamefont {Litvin}\ and\ \citenamefont
  {Opechowski}(1974)}]{ssgs4}%
  \BibitemOpen
  \bibfield  {author} {\bibinfo {author} {\bibfnamefont {D.B.}\ \bibnamefont
  {Litvin}}\ and\ \bibinfo {author} {\bibfnamefont {W.}~\bibnamefont
  {Opechowski}},\ }\bibfield  {title} {\enquote {\bibinfo {title} {Spin
  groups},}\ }\href {\doibase https://doi.org/10.1016/0031-8914(74)90157-8}
  {\bibfield  {journal} {\bibinfo  {journal} {Physica}\ }\textbf {\bibinfo
  {volume} {76}},\ \bibinfo {pages} {538--554} (\bibinfo {year}
  {1974})}\BibitemShut {NoStop}%
\bibitem [{\citenamefont {Corticelli}\ \emph {et~al.}(2022)\citenamefont
  {Corticelli}, \citenamefont {Moessner},\ and\ \citenamefont
  {McClarty}}]{ssgs5}%
  \BibitemOpen
  \bibfield  {author} {\bibinfo {author} {\bibfnamefont {A.}~\bibnamefont
  {Corticelli}}, \bibinfo {author} {\bibfnamefont {R.}~\bibnamefont
  {Moessner}}, \ and\ \bibinfo {author} {\bibfnamefont {P.~A.}\ \bibnamefont
  {McClarty}},\ }\bibfield  {title} {\enquote {\bibinfo {title} {Spin-space
  groups and magnon band topology},}\ }\href {\doibase
  10.1103/PhysRevB.105.064430} {\bibfield  {journal} {\bibinfo  {journal}
  {Phys. Rev. B}\ }\textbf {\bibinfo {volume} {105}},\ \bibinfo {pages}
  {064430} (\bibinfo {year} {2022})}\BibitemShut {NoStop}%
\bibitem [{\citenamefont {Schiff}\ \emph {et~al.}(2025)\citenamefont {Schiff},
  \citenamefont {Corticelli}, \citenamefont {Guerreiro}, \citenamefont
  {Romh{\'a}nyi},\ and\ \citenamefont {McClarty}}]{ssgs6}%
  \BibitemOpen
  \bibfield  {author} {\bibinfo {author} {\bibfnamefont {Hana}\ \bibnamefont
  {Schiff}}, \bibinfo {author} {\bibfnamefont {Alberto}\ \bibnamefont
  {Corticelli}}, \bibinfo {author} {\bibfnamefont {Afonso}\ \bibnamefont
  {Guerreiro}}, \bibinfo {author} {\bibfnamefont {Judit}\ \bibnamefont
  {Romh{\'a}nyi}}, \ and\ \bibinfo {author} {\bibfnamefont {Paul~A}\
  \bibnamefont {McClarty}},\ }\bibfield  {title} {\enquote {\bibinfo {title}
  {The crystallographic spin point groups and their representations},}\
  }\href@noop {} {\bibfield  {journal} {\bibinfo  {journal} {SciPost Physics}\
  }\textbf {\bibinfo {volume} {18}},\ \bibinfo {pages} {109} (\bibinfo {year}
  {2025})}\BibitemShut {NoStop}%
\bibitem [{\citenamefont {Parshukov}\ and\ \citenamefont
  {Schnyder}(2025)}]{parshukov2025}%
  \BibitemOpen
  \bibfield  {author} {\bibinfo {author} {\bibfnamefont {Kirill}\ \bibnamefont
  {Parshukov}}\ and\ \bibinfo {author} {\bibfnamefont {Andreas~P.}\
  \bibnamefont {Schnyder}},\ }\href {https://arxiv.org/abs/2507.10700}
  {\enquote {\bibinfo {title} {Exotic superconducting states in
  altermagnets},}\ } (\bibinfo {year} {2025}),\ \Eprint
  {http://arxiv.org/abs/2507.10700} {arXiv:2507.10700 [cond-mat.supr-con]}
  \BibitemShut {NoStop}%
\bibitem [{\citenamefont {Zhu}\ \emph {et~al.}(2023)\citenamefont {Zhu},
  \citenamefont {Zhuang}, \citenamefont {Wu},\ and\ \citenamefont
  {Yan}}]{Zhu2023}%
  \BibitemOpen
  \bibfield  {author} {\bibinfo {author} {\bibfnamefont {Di}~\bibnamefont
  {Zhu}}, \bibinfo {author} {\bibfnamefont {Zheng-Yang}\ \bibnamefont
  {Zhuang}}, \bibinfo {author} {\bibfnamefont {Zhigang}\ \bibnamefont {Wu}}, \
  and\ \bibinfo {author} {\bibfnamefont {Zhongbo}\ \bibnamefont {Yan}},\
  }\bibfield  {title} {\enquote {\bibinfo {title} {Topological
  superconductivity in two-dimensional altermagnetic metals},}\ }\href
  {\doibase 10.1103/PhysRevB.108.184505} {\bibfield  {journal} {\bibinfo
  {journal} {Phys. Rev. B}\ }\textbf {\bibinfo {volume} {108}},\ \bibinfo
  {pages} {184505} (\bibinfo {year} {2023})}\BibitemShut {NoStop}%
\bibitem [{\citenamefont {Brekke}\ \emph {et~al.}(2023)\citenamefont {Brekke},
  \citenamefont {Brataas},\ and\ \citenamefont {Sudb\o{}}}]{Sudbo2023}%
  \BibitemOpen
  \bibfield  {author} {\bibinfo {author} {\bibfnamefont {Bj\o{}rnulf}\
  \bibnamefont {Brekke}}, \bibinfo {author} {\bibfnamefont {Arne}\ \bibnamefont
  {Brataas}}, \ and\ \bibinfo {author} {\bibfnamefont {Asle}\ \bibnamefont
  {Sudb\o{}}},\ }\bibfield  {title} {\enquote {\bibinfo {title}
  {Two-dimensional altermagnets: Superconductivity in a minimal microscopic
  model},}\ }\href {\doibase 10.1103/PhysRevB.108.224421} {\bibfield  {journal}
  {\bibinfo  {journal} {Phys. Rev. B}\ }\textbf {\bibinfo {volume} {108}},\
  \bibinfo {pages} {224421} (\bibinfo {year} {2023})}\BibitemShut {NoStop}%
\bibitem [{\citenamefont {Heung}\ and\ \citenamefont
  {Franz}(2024)}]{Heung2024}%
  \BibitemOpen
  \bibfield  {author} {\bibinfo {author} {\bibfnamefont {Tsz~Fung}\
  \bibnamefont {Heung}}\ and\ \bibinfo {author} {\bibfnamefont {Marcel}\
  \bibnamefont {Franz}},\ }\href {https://arxiv.org/abs/2411.17964} {\enquote
  {\bibinfo {title} {Probing topological degeneracy on a torus using
  superconducting altermagnets},}\ } (\bibinfo {year} {2024}),\ \Eprint
  {http://arxiv.org/abs/2411.17964} {arXiv:2411.17964 [cond-mat.supr-con]}
  \BibitemShut {NoStop}%
\bibitem [{\citenamefont {de~Carvalho}\ and\ \citenamefont
  {Freire}(2024)}]{Carvalho2024}%
  \BibitemOpen
  \bibfield  {author} {\bibinfo {author} {\bibfnamefont {Vanuildo~S.}\
  \bibnamefont {de~Carvalho}}\ and\ \bibinfo {author} {\bibfnamefont {Hermann}\
  \bibnamefont {Freire}},\ }\bibfield  {title} {\enquote {\bibinfo {title}
  {Unconventional superconductivity in altermagnets with spin-orbit
  coupling},}\ }\href {\doibase 10.1103/PhysRevB.110.L220503} {\bibfield
  {journal} {\bibinfo  {journal} {Phys. Rev. B}\ }\textbf {\bibinfo {volume}
  {110}},\ \bibinfo {pages} {L220503} (\bibinfo {year} {2024})}\BibitemShut
  {NoStop}%
\bibitem [{\citenamefont {Leraand}\ \emph {et~al.}(2025)\citenamefont
  {Leraand}, \citenamefont {Maeland},\ and\ \citenamefont
  {Sudbø}}]{Leraand2025}%
  \BibitemOpen
  \bibfield  {author} {\bibinfo {author} {\bibfnamefont {Kristoffer}\
  \bibnamefont {Leraand}}, \bibinfo {author} {\bibfnamefont {Kristian}\
  \bibnamefont {Maeland}}, \ and\ \bibinfo {author} {\bibfnamefont {Asle}\
  \bibnamefont {Sudbø}},\ }\href {https://arxiv.org/abs/2502.08704} {\enquote
  {\bibinfo {title} {Phonon-mediated spin-polarized superconductivity in
  altermagnets},}\ } (\bibinfo {year} {2025}),\ \Eprint
  {http://arxiv.org/abs/2502.08704} {arXiv:2502.08704 [cond-mat.supr-con]}
  \BibitemShut {NoStop}%
\bibitem [{\citenamefont {Zhang}\ \emph
  {et~al.}(2024{\natexlab{a}})\citenamefont {Zhang}, \citenamefont {Cheng},
  \citenamefont {Yin}, \citenamefont {Liu}, \citenamefont {Deng}, \citenamefont
  {Qiao}, \citenamefont {Shi}, \citenamefont {Zhang}, \citenamefont {Lin},
  \citenamefont {Liu}, \citenamefont {Ye}, \citenamefont {Huang}, \citenamefont
  {Meng}, \citenamefont {Zhang}, \citenamefont {Okuda}, \citenamefont
  {Shimada}, \citenamefont {Cui}, \citenamefont {Zhao}, \citenamefont {Cao},
  \citenamefont {Qiao}, \citenamefont {Liu},\ and\ \citenamefont
  {Chen}}]{Zhang2024}%
  \BibitemOpen
  \bibfield  {author} {\bibinfo {author} {\bibfnamefont {Fayuan}\ \bibnamefont
  {Zhang}}, \bibinfo {author} {\bibfnamefont {Xingkai}\ \bibnamefont {Cheng}},
  \bibinfo {author} {\bibfnamefont {Zhouyi}\ \bibnamefont {Yin}}, \bibinfo
  {author} {\bibfnamefont {Changchao}\ \bibnamefont {Liu}}, \bibinfo {author}
  {\bibfnamefont {Liwei}\ \bibnamefont {Deng}}, \bibinfo {author}
  {\bibfnamefont {Yuxi}\ \bibnamefont {Qiao}}, \bibinfo {author} {\bibfnamefont
  {Zheng}\ \bibnamefont {Shi}}, \bibinfo {author} {\bibfnamefont {Shuxuan}\
  \bibnamefont {Zhang}}, \bibinfo {author} {\bibfnamefont {Junhao}\
  \bibnamefont {Lin}}, \bibinfo {author} {\bibfnamefont {Zhengtai}\
  \bibnamefont {Liu}}, \bibinfo {author} {\bibfnamefont {Mao}\ \bibnamefont
  {Ye}}, \bibinfo {author} {\bibfnamefont {Yaobo}\ \bibnamefont {Huang}},
  \bibinfo {author} {\bibfnamefont {Xiangyu}\ \bibnamefont {Meng}}, \bibinfo
  {author} {\bibfnamefont {Cheng}\ \bibnamefont {Zhang}}, \bibinfo {author}
  {\bibfnamefont {Taichi}\ \bibnamefont {Okuda}}, \bibinfo {author}
  {\bibfnamefont {Kenya}\ \bibnamefont {Shimada}}, \bibinfo {author}
  {\bibfnamefont {Shengtao}\ \bibnamefont {Cui}}, \bibinfo {author}
  {\bibfnamefont {Yue}\ \bibnamefont {Zhao}}, \bibinfo {author} {\bibfnamefont
  {Guang-Han}\ \bibnamefont {Cao}}, \bibinfo {author} {\bibfnamefont {Shan}\
  \bibnamefont {Qiao}}, \bibinfo {author} {\bibfnamefont {Junwei}\ \bibnamefont
  {Liu}}, \ and\ \bibinfo {author} {\bibfnamefont {Chaoyu}\ \bibnamefont
  {Chen}},\ }\href {https://arxiv.org/abs/2407.19555} {\enquote {\bibinfo
  {title} {Crystal-symmetry-paired spin-valley locking in a layered
  room-temperature antiferromagnet},}\ } (\bibinfo {year}
  {2024}{\natexlab{a}}),\ \Eprint {http://arxiv.org/abs/2407.19555}
  {arXiv:2407.19555 [cond-mat.str-el]} \BibitemShut {NoStop}%
\bibitem [{\citenamefont {Krempask{\'y}}\ \emph {et~al.}(2024)\citenamefont
  {Krempask{\'y}}, \citenamefont {{\v S}mejkal}, \citenamefont {D'Souza},
  \citenamefont {Hajlaoui}, \citenamefont {Springholz}, \citenamefont
  {Uhl{\'\i}{\v r}ov{\'a}}, \citenamefont {Alarab}, \citenamefont
  {Constantinou}, \citenamefont {Strocov}, \citenamefont {Usanov},
  \citenamefont {Pudelko}, \citenamefont {Gonz{\'a}lez-Hern{\'a}ndez},
  \citenamefont {Birk~Hellenes}, \citenamefont {Jansa}, \citenamefont
  {Reichlov{\'a}}, \citenamefont {{\v S}ob{\'a}{\v n}}, \citenamefont
  {Gonzalez~Betancourt}, \citenamefont {Wadley}, \citenamefont {Sinova},
  \citenamefont {Kriegner}, \citenamefont {Min{\'a}r}, \citenamefont {Dil},\
  and\ \citenamefont {Jungwirth}}]{Krempasky2024}%
  \BibitemOpen
  \bibfield  {author} {\bibinfo {author} {\bibfnamefont {J.}~\bibnamefont
  {Krempask{\'y}}}, \bibinfo {author} {\bibfnamefont {L.}~\bibnamefont {{\v
  S}mejkal}}, \bibinfo {author} {\bibfnamefont {S.~W.}\ \bibnamefont
  {D'Souza}}, \bibinfo {author} {\bibfnamefont {M.}~\bibnamefont {Hajlaoui}},
  \bibinfo {author} {\bibfnamefont {G.}~\bibnamefont {Springholz}}, \bibinfo
  {author} {\bibfnamefont {K.}~\bibnamefont {Uhl{\'\i}{\v r}ov{\'a}}}, \bibinfo
  {author} {\bibfnamefont {F.}~\bibnamefont {Alarab}}, \bibinfo {author}
  {\bibfnamefont {P.~C.}\ \bibnamefont {Constantinou}}, \bibinfo {author}
  {\bibfnamefont {V.}~\bibnamefont {Strocov}}, \bibinfo {author} {\bibfnamefont
  {D.}~\bibnamefont {Usanov}}, \bibinfo {author} {\bibfnamefont {W.~R.}\
  \bibnamefont {Pudelko}}, \bibinfo {author} {\bibfnamefont {R.}~\bibnamefont
  {Gonz{\'a}lez-Hern{\'a}ndez}}, \bibinfo {author} {\bibfnamefont
  {A.}~\bibnamefont {Birk~Hellenes}}, \bibinfo {author} {\bibfnamefont
  {Z.}~\bibnamefont {Jansa}}, \bibinfo {author} {\bibfnamefont
  {H.}~\bibnamefont {Reichlov{\'a}}}, \bibinfo {author} {\bibfnamefont
  {Z.}~\bibnamefont {{\v S}ob{\'a}{\v n}}}, \bibinfo {author} {\bibfnamefont
  {R.~D.}\ \bibnamefont {Gonzalez~Betancourt}}, \bibinfo {author}
  {\bibfnamefont {P.}~\bibnamefont {Wadley}}, \bibinfo {author} {\bibfnamefont
  {J.}~\bibnamefont {Sinova}}, \bibinfo {author} {\bibfnamefont
  {D.}~\bibnamefont {Kriegner}}, \bibinfo {author} {\bibfnamefont
  {J.}~\bibnamefont {Min{\'a}r}}, \bibinfo {author} {\bibfnamefont {J.~H.}\
  \bibnamefont {Dil}}, \ and\ \bibinfo {author} {\bibfnamefont
  {T.}~\bibnamefont {Jungwirth}},\ }\bibfield  {title} {\enquote {\bibinfo
  {title} {Altermagnetic lifting of kramers spin degeneracy},}\ }\href
  {\doibase 10.1038/s41586-023-06907-7} {\bibfield  {journal} {\bibinfo
  {journal} {Nature}\ }\textbf {\bibinfo {volume} {626}},\ \bibinfo {pages}
  {517--522} (\bibinfo {year} {2024})}\BibitemShut {NoStop}%
\bibitem [{\citenamefont {Lee}\ \emph {et~al.}(2024)\citenamefont {Lee},
  \citenamefont {Lee}, \citenamefont {Jung}, \citenamefont {Jung},
  \citenamefont {Kim}, \citenamefont {Lee}, \citenamefont {Seok}, \citenamefont
  {Kim}, \citenamefont {Park}, \citenamefont {\ifmmode~\check{S}\else
  \v{S}\fi{}mejkal}, \citenamefont {Kang},\ and\ \citenamefont
  {Kim}}]{Lee2024}%
  \BibitemOpen
  \bibfield  {author} {\bibinfo {author} {\bibfnamefont {Suyoung}\ \bibnamefont
  {Lee}}, \bibinfo {author} {\bibfnamefont {Sangjae}\ \bibnamefont {Lee}},
  \bibinfo {author} {\bibfnamefont {Saegyeol}\ \bibnamefont {Jung}}, \bibinfo
  {author} {\bibfnamefont {Jiwon}\ \bibnamefont {Jung}}, \bibinfo {author}
  {\bibfnamefont {Donghan}\ \bibnamefont {Kim}}, \bibinfo {author}
  {\bibfnamefont {Yeonjae}\ \bibnamefont {Lee}}, \bibinfo {author}
  {\bibfnamefont {Byeongjun}\ \bibnamefont {Seok}}, \bibinfo {author}
  {\bibfnamefont {Jaeyoung}\ \bibnamefont {Kim}}, \bibinfo {author}
  {\bibfnamefont {Byeong~Gyu}\ \bibnamefont {Park}}, \bibinfo {author}
  {\bibfnamefont {Libor}\ \bibnamefont {\ifmmode~\check{S}\else
  \v{S}\fi{}mejkal}}, \bibinfo {author} {\bibfnamefont {Chang-Jong}\
  \bibnamefont {Kang}}, \ and\ \bibinfo {author} {\bibfnamefont {Changyoung}\
  \bibnamefont {Kim}},\ }\bibfield  {title} {\enquote {\bibinfo {title} {Broken
  kramers degeneracy in altermagnetic mnte},}\ }\href {\doibase
  10.1103/PhysRevLett.132.036702} {\bibfield  {journal} {\bibinfo  {journal}
  {Phys. Rev. Lett.}\ }\textbf {\bibinfo {volume} {132}},\ \bibinfo {pages}
  {036702} (\bibinfo {year} {2024})}\BibitemShut {NoStop}%
\bibitem [{\citenamefont {Fedchenko}\ \emph {et~al.}(2024)\citenamefont
  {Fedchenko}, \citenamefont {Min\'ar}, \citenamefont {Akashdeep},
  \citenamefont {D’Souza}, \citenamefont {Vasilyev}, \citenamefont {Tkach},
  \citenamefont {Odenbreit}, \citenamefont {Nguyen}, \citenamefont
  {Kutnyakhov}, \citenamefont {Wind}, \citenamefont {Wenthaus}, \citenamefont
  {Scholz}, \citenamefont {Rossnagel}, \citenamefont {Hoesch}, \citenamefont
  {Aeschlimann}, \citenamefont {Stadtmüller}, \citenamefont {Kläui},
  \citenamefont {Schönhense}, \citenamefont {Jungwirth}, \citenamefont
  {Hellenes}, \citenamefont {Jakob}, \citenamefont {\ifmmode~\check{S}\else
  \v{S}\fi{}mejkal}, \citenamefont {Sinova},\ and\ \citenamefont
  {Elmers}}]{Fedchenko2024}%
  \BibitemOpen
  \bibfield  {author} {\bibinfo {author} {\bibfnamefont {Olena}\ \bibnamefont
  {Fedchenko}}, \bibinfo {author} {\bibfnamefont {Jan}\ \bibnamefont
  {Min\'ar}}, \bibinfo {author} {\bibfnamefont {Akashdeep}\ \bibnamefont
  {Akashdeep}}, \bibinfo {author} {\bibfnamefont {Sunil~Wilfred}\ \bibnamefont
  {D’Souza}}, \bibinfo {author} {\bibfnamefont {Dmitry}\ \bibnamefont
  {Vasilyev}}, \bibinfo {author} {\bibfnamefont {Olena}\ \bibnamefont {Tkach}},
  \bibinfo {author} {\bibfnamefont {Lukas}\ \bibnamefont {Odenbreit}}, \bibinfo
  {author} {\bibfnamefont {Quynh}\ \bibnamefont {Nguyen}}, \bibinfo {author}
  {\bibfnamefont {Dmytro}\ \bibnamefont {Kutnyakhov}}, \bibinfo {author}
  {\bibfnamefont {Nils}\ \bibnamefont {Wind}}, \bibinfo {author} {\bibfnamefont
  {Lukas}\ \bibnamefont {Wenthaus}}, \bibinfo {author} {\bibfnamefont {Markus}\
  \bibnamefont {Scholz}}, \bibinfo {author} {\bibfnamefont {Kai}\ \bibnamefont
  {Rossnagel}}, \bibinfo {author} {\bibfnamefont {Moritz}\ \bibnamefont
  {Hoesch}}, \bibinfo {author} {\bibfnamefont {Martin}\ \bibnamefont
  {Aeschlimann}}, \bibinfo {author} {\bibfnamefont {Benjamin}\ \bibnamefont
  {Stadtmüller}}, \bibinfo {author} {\bibfnamefont {Mathias}\ \bibnamefont
  {Kläui}}, \bibinfo {author} {\bibfnamefont {Gerd}\ \bibnamefont
  {Schönhense}}, \bibinfo {author} {\bibfnamefont {Tomas}\ \bibnamefont
  {Jungwirth}}, \bibinfo {author} {\bibfnamefont {Anna~Birk}\ \bibnamefont
  {Hellenes}}, \bibinfo {author} {\bibfnamefont {Gerhard}\ \bibnamefont
  {Jakob}}, \bibinfo {author} {\bibfnamefont {Libor}\ \bibnamefont
  {\ifmmode~\check{S}\else \v{S}\fi{}mejkal}}, \bibinfo {author} {\bibfnamefont
  {Jairo}\ \bibnamefont {Sinova}}, \ and\ \bibinfo {author} {\bibfnamefont
  {Hans-Joachim}\ \bibnamefont {Elmers}},\ }\bibfield  {title} {\enquote
  {\bibinfo {title} {Observation of time-reversal symmetry breaking in the band
  structure of altermagnetic ruo<sub>2</sub>},}\ }\href {\doibase
  10.1126/sciadv.adj4883} {\bibfield  {journal} {\bibinfo  {journal} {Science
  Advances}\ }\textbf {\bibinfo {volume} {10}},\ \bibinfo {pages} {eadj4883}
  (\bibinfo {year} {2024})}\BibitemShut {NoStop}%
\bibitem [{\citenamefont {Reimers}\ \emph {et~al.}(2024)\citenamefont
  {Reimers}, \citenamefont {Odenbreit}, \citenamefont {{\v S}mejkal},
  \citenamefont {Strocov}, \citenamefont {Constantinou}, \citenamefont
  {Hellenes}, \citenamefont {Jaeschke~Ubiergo}, \citenamefont {Campos},
  \citenamefont {Bharadwaj}, \citenamefont {Chakraborty}, \citenamefont
  {Denneulin}, \citenamefont {Shi}, \citenamefont {Dunin-Borkowski},
  \citenamefont {Das}, \citenamefont {Kl{\"a}ui}, \citenamefont {Sinova},\ and\
  \citenamefont {Jourdan}}]{Reimers2024}%
  \BibitemOpen
  \bibfield  {author} {\bibinfo {author} {\bibfnamefont {Sonka}\ \bibnamefont
  {Reimers}}, \bibinfo {author} {\bibfnamefont {Lukas}\ \bibnamefont
  {Odenbreit}}, \bibinfo {author} {\bibfnamefont {Libor}\ \bibnamefont {{\v
  S}mejkal}}, \bibinfo {author} {\bibfnamefont {Vladimir~N.}\ \bibnamefont
  {Strocov}}, \bibinfo {author} {\bibfnamefont {Procopios}\ \bibnamefont
  {Constantinou}}, \bibinfo {author} {\bibfnamefont {Anna~B.}\ \bibnamefont
  {Hellenes}}, \bibinfo {author} {\bibfnamefont {Rodrigo}\ \bibnamefont
  {Jaeschke~Ubiergo}}, \bibinfo {author} {\bibfnamefont {Warlley~H.}\
  \bibnamefont {Campos}}, \bibinfo {author} {\bibfnamefont {Venkata~K.}\
  \bibnamefont {Bharadwaj}}, \bibinfo {author} {\bibfnamefont {Atasi}\
  \bibnamefont {Chakraborty}}, \bibinfo {author} {\bibfnamefont {Thibaud}\
  \bibnamefont {Denneulin}}, \bibinfo {author} {\bibfnamefont {Wen}\
  \bibnamefont {Shi}}, \bibinfo {author} {\bibfnamefont {Rafal~E.}\
  \bibnamefont {Dunin-Borkowski}}, \bibinfo {author} {\bibfnamefont {Suvadip}\
  \bibnamefont {Das}}, \bibinfo {author} {\bibfnamefont {Mathias}\ \bibnamefont
  {Kl{\"a}ui}}, \bibinfo {author} {\bibfnamefont {Jairo}\ \bibnamefont
  {Sinova}}, \ and\ \bibinfo {author} {\bibfnamefont {Martin}\ \bibnamefont
  {Jourdan}},\ }\bibfield  {title} {\enquote {\bibinfo {title} {Direct
  observation of altermagnetic band splitting in crsb thin films},}\ }\href
  {\doibase 10.1038/s41467-024-46476-5} {\bibfield  {journal} {\bibinfo
  {journal} {Nature Communications}\ }\textbf {\bibinfo {volume} {15}},\
  \bibinfo {pages} {2116} (\bibinfo {year} {2024})}\BibitemShut {NoStop}%
\bibitem [{\citenamefont {Ding}\ \emph {et~al.}(2024)\citenamefont {Ding},
  \citenamefont {Jiang}, \citenamefont {Chen}, \citenamefont {Tao},
  \citenamefont {Liu}, \citenamefont {Li}, \citenamefont {Liu}, \citenamefont
  {Sun}, \citenamefont {Cheng}, \citenamefont {Liu}, \citenamefont {Yang},
  \citenamefont {Zhang}, \citenamefont {Deng}, \citenamefont {Jing},
  \citenamefont {Huang}, \citenamefont {Shi}, \citenamefont {Ye}, \citenamefont
  {Qiao}, \citenamefont {Wang}, \citenamefont {Guo}, \citenamefont {Feng},\
  and\ \citenamefont {Shen}}]{Ding2024}%
  \BibitemOpen
  \bibfield  {author} {\bibinfo {author} {\bibfnamefont {Jianyang}\
  \bibnamefont {Ding}}, \bibinfo {author} {\bibfnamefont {Zhicheng}\
  \bibnamefont {Jiang}}, \bibinfo {author} {\bibfnamefont {Xiuhua}\
  \bibnamefont {Chen}}, \bibinfo {author} {\bibfnamefont {Zicheng}\
  \bibnamefont {Tao}}, \bibinfo {author} {\bibfnamefont {Zhengtai}\
  \bibnamefont {Liu}}, \bibinfo {author} {\bibfnamefont {Tongrui}\ \bibnamefont
  {Li}}, \bibinfo {author} {\bibfnamefont {Jishan}\ \bibnamefont {Liu}},
  \bibinfo {author} {\bibfnamefont {Jianping}\ \bibnamefont {Sun}}, \bibinfo
  {author} {\bibfnamefont {Jinguang}\ \bibnamefont {Cheng}}, \bibinfo {author}
  {\bibfnamefont {Jiayu}\ \bibnamefont {Liu}}, \bibinfo {author} {\bibfnamefont
  {Yichen}\ \bibnamefont {Yang}}, \bibinfo {author} {\bibfnamefont {Runfeng}\
  \bibnamefont {Zhang}}, \bibinfo {author} {\bibfnamefont {Liwei}\ \bibnamefont
  {Deng}}, \bibinfo {author} {\bibfnamefont {Wenchuan}\ \bibnamefont {Jing}},
  \bibinfo {author} {\bibfnamefont {Yu}~\bibnamefont {Huang}}, \bibinfo
  {author} {\bibfnamefont {Yuming}\ \bibnamefont {Shi}}, \bibinfo {author}
  {\bibfnamefont {Mao}\ \bibnamefont {Ye}}, \bibinfo {author} {\bibfnamefont
  {Shan}\ \bibnamefont {Qiao}}, \bibinfo {author} {\bibfnamefont {Yilin}\
  \bibnamefont {Wang}}, \bibinfo {author} {\bibfnamefont {Yanfeng}\
  \bibnamefont {Guo}}, \bibinfo {author} {\bibfnamefont {Donglai}\ \bibnamefont
  {Feng}}, \ and\ \bibinfo {author} {\bibfnamefont {Dawei}\ \bibnamefont
  {Shen}},\ }\bibfield  {title} {\enquote {\bibinfo {title} {Large band
  splitting in $g$-wave altermagnet crsb},}\ }\href {\doibase
  10.1103/PhysRevLett.133.206401} {\bibfield  {journal} {\bibinfo  {journal}
  {Phys. Rev. Lett.}\ }\textbf {\bibinfo {volume} {133}},\ \bibinfo {pages}
  {206401} (\bibinfo {year} {2024})}\BibitemShut {NoStop}%
\bibitem [{\citenamefont {Yang}\ \emph {et~al.}(2025)\citenamefont {Yang},
  \citenamefont {Li}, \citenamefont {Yang}, \citenamefont {Li}, \citenamefont
  {Zheng}, \citenamefont {Zhu}, \citenamefont {Pan}, \citenamefont {Xu},
  \citenamefont {Cao}, \citenamefont {Zhao}, \citenamefont {Jana},
  \citenamefont {Zhang}, \citenamefont {Ye}, \citenamefont {Song},
  \citenamefont {Hu}, \citenamefont {Yang}, \citenamefont {Fujii},
  \citenamefont {Vobornik}, \citenamefont {Shi}, \citenamefont {Yuan},
  \citenamefont {Zhang}, \citenamefont {Xu},\ and\ \citenamefont
  {Liu}}]{Yang2025}%
  \BibitemOpen
  \bibfield  {author} {\bibinfo {author} {\bibfnamefont {Guowei}\ \bibnamefont
  {Yang}}, \bibinfo {author} {\bibfnamefont {Zhanghuan}\ \bibnamefont {Li}},
  \bibinfo {author} {\bibfnamefont {Sai}\ \bibnamefont {Yang}}, \bibinfo
  {author} {\bibfnamefont {Jiyuan}\ \bibnamefont {Li}}, \bibinfo {author}
  {\bibfnamefont {Hao}\ \bibnamefont {Zheng}}, \bibinfo {author} {\bibfnamefont
  {Weifan}\ \bibnamefont {Zhu}}, \bibinfo {author} {\bibfnamefont
  {Ze}~\bibnamefont {Pan}}, \bibinfo {author} {\bibfnamefont {Yifu}\
  \bibnamefont {Xu}}, \bibinfo {author} {\bibfnamefont {Saizheng}\ \bibnamefont
  {Cao}}, \bibinfo {author} {\bibfnamefont {Wenxuan}\ \bibnamefont {Zhao}},
  \bibinfo {author} {\bibfnamefont {Anupam}\ \bibnamefont {Jana}}, \bibinfo
  {author} {\bibfnamefont {Jiawen}\ \bibnamefont {Zhang}}, \bibinfo {author}
  {\bibfnamefont {Mao}\ \bibnamefont {Ye}}, \bibinfo {author} {\bibfnamefont
  {Yu}~\bibnamefont {Song}}, \bibinfo {author} {\bibfnamefont {Lun-Hui}\
  \bibnamefont {Hu}}, \bibinfo {author} {\bibfnamefont {Lexian}\ \bibnamefont
  {Yang}}, \bibinfo {author} {\bibfnamefont {Jun}\ \bibnamefont {Fujii}},
  \bibinfo {author} {\bibfnamefont {Ivana}\ \bibnamefont {Vobornik}}, \bibinfo
  {author} {\bibfnamefont {Ming}\ \bibnamefont {Shi}}, \bibinfo {author}
  {\bibfnamefont {Huiqiu}\ \bibnamefont {Yuan}}, \bibinfo {author}
  {\bibfnamefont {Yongjun}\ \bibnamefont {Zhang}}, \bibinfo {author}
  {\bibfnamefont {Yuanfeng}\ \bibnamefont {Xu}}, \ and\ \bibinfo {author}
  {\bibfnamefont {Yang}\ \bibnamefont {Liu}},\ }\href
  {https://arxiv.org/abs/2405.12575} {\enquote {\bibinfo {title}
  {Three-dimensional mapping of the altermagnetic spin splitting in crsb},}\ }
  (\bibinfo {year} {2025}),\ \Eprint {http://arxiv.org/abs/2405.12575}
  {arXiv:2405.12575 [cond-mat.mtrl-sci]} \BibitemShut {NoStop}%
\bibitem [{\citenamefont {Naka}\ \emph {et~al.}(2019)\citenamefont {Naka},
  \citenamefont {Hayami}, \citenamefont {Kusunose}, \citenamefont {Yanagi},
  \citenamefont {Motome},\ and\ \citenamefont {Seo}}]{Naka2019}%
  \BibitemOpen
  \bibfield  {author} {\bibinfo {author} {\bibfnamefont {Makoto}\ \bibnamefont
  {Naka}}, \bibinfo {author} {\bibfnamefont {Satoru}\ \bibnamefont {Hayami}},
  \bibinfo {author} {\bibfnamefont {Hiroaki}\ \bibnamefont {Kusunose}},
  \bibinfo {author} {\bibfnamefont {Yuki}\ \bibnamefont {Yanagi}}, \bibinfo
  {author} {\bibfnamefont {Yukitoshi}\ \bibnamefont {Motome}}, \ and\ \bibinfo
  {author} {\bibfnamefont {Hitoshi}\ \bibnamefont {Seo}},\ }\bibfield  {title}
  {\enquote {\bibinfo {title} {Spin current generation in organic
  antiferromagnets},}\ }\href {\doibase 10.1038/s41467-019-12229-y} {\bibfield
  {journal} {\bibinfo  {journal} {Nature Communications}\ }\textbf {\bibinfo
  {volume} {10}},\ \bibinfo {pages} {4305} (\bibinfo {year}
  {2019})}\BibitemShut {NoStop}%
\bibitem [{\citenamefont {Yuan}\ \emph {et~al.}(2020)\citenamefont {Yuan},
  \citenamefont {Wang}, \citenamefont {Luo}, \citenamefont {Rashba},\ and\
  \citenamefont {Zunger}}]{Yuan2020}%
  \BibitemOpen
  \bibfield  {author} {\bibinfo {author} {\bibfnamefont {Lin-Ding}\
  \bibnamefont {Yuan}}, \bibinfo {author} {\bibfnamefont {Zhi}\ \bibnamefont
  {Wang}}, \bibinfo {author} {\bibfnamefont {Jun-Wei}\ \bibnamefont {Luo}},
  \bibinfo {author} {\bibfnamefont {Emmanuel~I.}\ \bibnamefont {Rashba}}, \
  and\ \bibinfo {author} {\bibfnamefont {Alex}\ \bibnamefont {Zunger}},\
  }\bibfield  {title} {\enquote {\bibinfo {title} {Giant momentum-dependent
  spin splitting in centrosymmetric low-$z$ antiferromagnets},}\ }\href
  {\doibase 10.1103/PhysRevB.102.014422} {\bibfield  {journal} {\bibinfo
  {journal} {Phys. Rev. B}\ }\textbf {\bibinfo {volume} {102}},\ \bibinfo
  {pages} {014422} (\bibinfo {year} {2020})}\BibitemShut {NoStop}%
\bibitem [{\citenamefont {Mazin}\ \emph {et~al.}(2021)\citenamefont {Mazin},
  \citenamefont {Koepernik}, \citenamefont {Johannes}, \citenamefont
  {González-Hernández},\ and\ \citenamefont {Šmejkal}}]{Mazin2021}%
  \BibitemOpen
  \bibfield  {author} {\bibinfo {author} {\bibfnamefont {Igor~I.}\ \bibnamefont
  {Mazin}}, \bibinfo {author} {\bibfnamefont {Klaus}\ \bibnamefont
  {Koepernik}}, \bibinfo {author} {\bibfnamefont {Michelle~D.}\ \bibnamefont
  {Johannes}}, \bibinfo {author} {\bibfnamefont {Rafael}\ \bibnamefont
  {González-Hernández}}, \ and\ \bibinfo {author} {\bibfnamefont {Libor}\
  \bibnamefont {Šmejkal}},\ }\bibfield  {title} {\enquote {\bibinfo {title}
  {Prediction of unconventional magnetism in doped fesb<sub>2</sub>},}\ }\href
  {\doibase 10.1073/pnas.2108924118} {\bibfield  {journal} {\bibinfo  {journal}
  {Proceedings of the National Academy of Sciences}\ }\textbf {\bibinfo
  {volume} {118}},\ \bibinfo {pages} {e2108924118} (\bibinfo {year}
  {2021})}\BibitemShut {NoStop}%
\bibitem [{\citenamefont {Bhowal}\ and\ \citenamefont
  {Spaldin}(2022)}]{Spaldin2022}%
  \BibitemOpen
  \bibfield  {author} {\bibinfo {author} {\bibfnamefont {Sayantika}\
  \bibnamefont {Bhowal}}\ and\ \bibinfo {author} {\bibfnamefont {Nicola~A.}\
  \bibnamefont {Spaldin}},\ }\href {https://arxiv.org/abs/2212.03756} {\enquote
  {\bibinfo {title} {Magnetic octupoles as the order parameter for
  unconventional antiferromagnetism},}\ } (\bibinfo {year} {2022}),\ \Eprint
  {http://arxiv.org/abs/2212.03756} {arXiv:2212.03756 [cond-mat.str-el]}
  \BibitemShut {NoStop}%
\bibitem [{\citenamefont {Guo}\ \emph {et~al.}(2023)\citenamefont {Guo},
  \citenamefont {Liu}, \citenamefont {Janson}, \citenamefont {Fulga},
  \citenamefont {{van den Brink}},\ and\ \citenamefont {Facio}}]{Guo2023}%
  \BibitemOpen
  \bibfield  {author} {\bibinfo {author} {\bibfnamefont {Yaqian}\ \bibnamefont
  {Guo}}, \bibinfo {author} {\bibfnamefont {Hui}\ \bibnamefont {Liu}}, \bibinfo
  {author} {\bibfnamefont {Oleg}\ \bibnamefont {Janson}}, \bibinfo {author}
  {\bibfnamefont {Ion~Cosma}\ \bibnamefont {Fulga}}, \bibinfo {author}
  {\bibfnamefont {Jeroen}\ \bibnamefont {{van den Brink}}}, \ and\ \bibinfo
  {author} {\bibfnamefont {Jorge~I.}\ \bibnamefont {Facio}},\ }\bibfield
  {title} {\enquote {\bibinfo {title} {Spin-split collinear antiferromagnets: A
  large-scale ab-initio study},}\ }\href {\doibase
  https://doi.org/10.1016/j.mtphys.2023.100991} {\bibfield  {journal} {\bibinfo
   {journal} {Materials Today Physics}\ }\textbf {\bibinfo {volume} {32}},\
  \bibinfo {pages} {100991} (\bibinfo {year} {2023})}\BibitemShut {NoStop}%
\bibitem [{\citenamefont {Maznichenko}\ \emph {et~al.}(2024)\citenamefont
  {Maznichenko}, \citenamefont {Ernst}, \citenamefont {Maryenko}, \citenamefont
  {Dugaev}, \citenamefont {Sherman}, \citenamefont {Buczek}, \citenamefont
  {Parkin},\ and\ \citenamefont {Ostanin}}]{Ostanin2024}%
  \BibitemOpen
  \bibfield  {author} {\bibinfo {author} {\bibfnamefont {I.~V.}\ \bibnamefont
  {Maznichenko}}, \bibinfo {author} {\bibfnamefont {A.}~\bibnamefont {Ernst}},
  \bibinfo {author} {\bibfnamefont {D.}~\bibnamefont {Maryenko}}, \bibinfo
  {author} {\bibfnamefont {V.~K.}\ \bibnamefont {Dugaev}}, \bibinfo {author}
  {\bibfnamefont {E.~Ya.}\ \bibnamefont {Sherman}}, \bibinfo {author}
  {\bibfnamefont {P.}~\bibnamefont {Buczek}}, \bibinfo {author} {\bibfnamefont
  {S.~S.~P.}\ \bibnamefont {Parkin}}, \ and\ \bibinfo {author} {\bibfnamefont
  {S.}~\bibnamefont {Ostanin}},\ }\bibfield  {title} {\enquote {\bibinfo
  {title} {Fragile altermagnetism and orbital disorder in mott insulator
  ${\mathrm{latio}}_{3}$},}\ }\href {\doibase
  10.1103/PhysRevMaterials.8.064403} {\bibfield  {journal} {\bibinfo  {journal}
  {Phys. Rev. Mater.}\ }\textbf {\bibinfo {volume} {8}},\ \bibinfo {pages}
  {064403} (\bibinfo {year} {2024})}\BibitemShut {NoStop}%
\bibitem [{\citenamefont {Liu}\ \emph {et~al.}(2024)\citenamefont {Liu},
  \citenamefont {Yu},\ and\ \citenamefont {Liu}}]{Liu2024b}%
  \BibitemOpen
  \bibfield  {author} {\bibinfo {author} {\bibfnamefont {Yichen}\ \bibnamefont
  {Liu}}, \bibinfo {author} {\bibfnamefont {Junxi}\ \bibnamefont {Yu}}, \ and\
  \bibinfo {author} {\bibfnamefont {Cheng-Cheng}\ \bibnamefont {Liu}},\
  }\bibfield  {title} {\enquote {\bibinfo {title} {Twisted magnetic van der
  waals bilayers: An ideal platform for altermagnetism},}\ }\href {\doibase
  10.1103/PhysRevLett.133.206702} {\bibfield  {journal} {\bibinfo  {journal}
  {Phys. Rev. Lett.}\ }\textbf {\bibinfo {volume} {133}},\ \bibinfo {pages}
  {206702} (\bibinfo {year} {2024})}\BibitemShut {NoStop}%
\bibitem [{\citenamefont {Ouassou}\ \emph {et~al.}(2023)\citenamefont
  {Ouassou}, \citenamefont {Brataas},\ and\ \citenamefont
  {Linder}}]{Linder2023}%
  \BibitemOpen
  \bibfield  {author} {\bibinfo {author} {\bibfnamefont {Jabir~Ali}\
  \bibnamefont {Ouassou}}, \bibinfo {author} {\bibfnamefont {Arne}\
  \bibnamefont {Brataas}}, \ and\ \bibinfo {author} {\bibfnamefont {Jacob}\
  \bibnamefont {Linder}},\ }\bibfield  {title} {\enquote {\bibinfo {title} {dc
  josephson effect in altermagnets},}\ }\href {\doibase
  10.1103/PhysRevLett.131.076003} {\bibfield  {journal} {\bibinfo  {journal}
  {Phys. Rev. Lett.}\ }\textbf {\bibinfo {volume} {131}},\ \bibinfo {pages}
  {076003} (\bibinfo {year} {2023})}\BibitemShut {NoStop}%
\bibitem [{\citenamefont {Papaj}(2023)}]{Papaj2023}%
  \BibitemOpen
  \bibfield  {author} {\bibinfo {author} {\bibfnamefont {Micha\l{}}\
  \bibnamefont {Papaj}},\ }\bibfield  {title} {\enquote {\bibinfo {title}
  {Andreev reflection at the altermagnet-superconductor interface},}\ }\href
  {\doibase 10.1103/PhysRevB.108.L060508} {\bibfield  {journal} {\bibinfo
  {journal} {Phys. Rev. B}\ }\textbf {\bibinfo {volume} {108}},\ \bibinfo
  {pages} {L060508} (\bibinfo {year} {2023})}\BibitemShut {NoStop}%
\bibitem [{\citenamefont {Beenakker}\ and\ \citenamefont
  {Vakhtel}(2023)}]{Beenakker2023}%
  \BibitemOpen
  \bibfield  {author} {\bibinfo {author} {\bibfnamefont {C.~W.~J.}\
  \bibnamefont {Beenakker}}\ and\ \bibinfo {author} {\bibfnamefont
  {T.}~\bibnamefont {Vakhtel}},\ }\bibfield  {title} {\enquote {\bibinfo
  {title} {Phase-shifted andreev levels in an altermagnet josephson
  junction},}\ }\href {\doibase 10.1103/PhysRevB.108.075425} {\bibfield
  {journal} {\bibinfo  {journal} {Phys. Rev. B}\ }\textbf {\bibinfo {volume}
  {108}},\ \bibinfo {pages} {075425} (\bibinfo {year} {2023})}\BibitemShut
  {NoStop}%
\bibitem [{\citenamefont {Giil}\ \emph {et~al.}(2024)\citenamefont {Giil},
  \citenamefont {Brekke}, \citenamefont {Linder},\ and\ \citenamefont
  {Brataas}}]{Brataas2024}%
  \BibitemOpen
  \bibfield  {author} {\bibinfo {author} {\bibfnamefont {Hans~Gl\o{}ckner}\
  \bibnamefont {Giil}}, \bibinfo {author} {\bibfnamefont {Bj\o{}rnulf}\
  \bibnamefont {Brekke}}, \bibinfo {author} {\bibfnamefont {Jacob}\
  \bibnamefont {Linder}}, \ and\ \bibinfo {author} {\bibfnamefont {Arne}\
  \bibnamefont {Brataas}},\ }\bibfield  {title} {\enquote {\bibinfo {title}
  {Quasiclassical theory of superconducting spin-splitter effects and
  spin-filtering via altermagnets},}\ }\href {\doibase
  10.1103/PhysRevB.110.L140506} {\bibfield  {journal} {\bibinfo  {journal}
  {Phys. Rev. B}\ }\textbf {\bibinfo {volume} {110}},\ \bibinfo {pages}
  {L140506} (\bibinfo {year} {2024})}\BibitemShut {NoStop}%
\bibitem [{\citenamefont {Zhang}\ \emph
  {et~al.}(2024{\natexlab{b}})\citenamefont {Zhang}, \citenamefont {Hu},\ and\
  \citenamefont {Neupert}}]{Neupert2024}%
  \BibitemOpen
  \bibfield  {author} {\bibinfo {author} {\bibfnamefont {Song-Bo}\ \bibnamefont
  {Zhang}}, \bibinfo {author} {\bibfnamefont {Lun-Hui}\ \bibnamefont {Hu}}, \
  and\ \bibinfo {author} {\bibfnamefont {Titus}\ \bibnamefont {Neupert}},\
  }\bibfield  {title} {\enquote {\bibinfo {title} {Finite-momentum cooper
  pairing in proximitized altermagnets},}\ }\href {\doibase
  10.1038/s41467-024-45951-3} {\bibfield  {journal} {\bibinfo  {journal}
  {Nature Communications}\ }\textbf {\bibinfo {volume} {15}},\ \bibinfo {pages}
  {1801} (\bibinfo {year} {2024}{\natexlab{b}})}\BibitemShut {NoStop}%
\bibitem [{\citenamefont {Banerjee}\ and\ \citenamefont
  {Scheurer}(2024)}]{Banerjee2024}%
  \BibitemOpen
  \bibfield  {author} {\bibinfo {author} {\bibfnamefont {Sayan}\ \bibnamefont
  {Banerjee}}\ and\ \bibinfo {author} {\bibfnamefont {Mathias~S.}\ \bibnamefont
  {Scheurer}},\ }\bibfield  {title} {\enquote {\bibinfo {title} {Altermagnetic
  superconducting diode effect},}\ }\href {\doibase
  10.1103/PhysRevB.110.024503} {\bibfield  {journal} {\bibinfo  {journal}
  {Phys. Rev. B}\ }\textbf {\bibinfo {volume} {110}},\ \bibinfo {pages}
  {024503} (\bibinfo {year} {2024})}\BibitemShut {NoStop}%
\bibitem [{\citenamefont {Roig}\ \emph {et~al.}(2024)\citenamefont {Roig},
  \citenamefont {Kreisel}, \citenamefont {Yu}, \citenamefont {Andersen},\ and\
  \citenamefont {Agterberg}}]{sitedecoupling1}%
  \BibitemOpen
  \bibfield  {author} {\bibinfo {author} {\bibfnamefont {Merc\`e}\ \bibnamefont
  {Roig}}, \bibinfo {author} {\bibfnamefont {Andreas}\ \bibnamefont {Kreisel}},
  \bibinfo {author} {\bibfnamefont {Yue}\ \bibnamefont {Yu}}, \bibinfo {author}
  {\bibfnamefont {Brian~M.}\ \bibnamefont {Andersen}}, \ and\ \bibinfo {author}
  {\bibfnamefont {Daniel~F.}\ \bibnamefont {Agterberg}},\ }\bibfield  {title}
  {\enquote {\bibinfo {title} {Minimal models for altermagnetism},}\ }\href
  {\doibase 10.1103/PhysRevB.110.144412} {\bibfield  {journal} {\bibinfo
  {journal} {Phys. Rev. B}\ }\textbf {\bibinfo {volume} {110}},\ \bibinfo
  {pages} {144412} (\bibinfo {year} {2024})}\BibitemShut {NoStop}%
\bibitem [{\citenamefont {Heinsdorf}(2025)}]{sitedecoupling2}%
  \BibitemOpen
  \bibfield  {author} {\bibinfo {author} {\bibfnamefont {Niclas}\ \bibnamefont
  {Heinsdorf}},\ }\bibfield  {title} {\enquote {\bibinfo {title} {Altermagnetic
  instabilities from quantum geometry},}\ }\href {\doibase
  10.1103/PhysRevB.111.174407} {\bibfield  {journal} {\bibinfo  {journal}
  {Phys. Rev. B}\ }\textbf {\bibinfo {volume} {111}},\ \bibinfo {pages}
  {174407} (\bibinfo {year} {2025})}\BibitemShut {NoStop}%
\bibitem [{\citenamefont {Zhu}\ \emph {et~al.}(2025)\citenamefont {Zhu},
  \citenamefont {Huo}, \citenamefont {Feng}, \citenamefont {Zhang},
  \citenamefont {Yang},\ and\ \citenamefont {Guo}}]{Guo2025}%
  \BibitemOpen
  \bibfield  {author} {\bibinfo {author} {\bibfnamefont {Xingchuan}\
  \bibnamefont {Zhu}}, \bibinfo {author} {\bibfnamefont {Xingmin}\ \bibnamefont
  {Huo}}, \bibinfo {author} {\bibfnamefont {Shiping}\ \bibnamefont {Feng}},
  \bibinfo {author} {\bibfnamefont {Song-Bo}\ \bibnamefont {Zhang}}, \bibinfo
  {author} {\bibfnamefont {Shengyuan~A.}\ \bibnamefont {Yang}}, \ and\ \bibinfo
  {author} {\bibfnamefont {Huaiming}\ \bibnamefont {Guo}},\ }\bibfield  {title}
  {\enquote {\bibinfo {title} {Design of altermagnetic models from spin
  clusters},}\ }\href {\doibase 10.1103/PhysRevLett.134.166701} {\bibfield
  {journal} {\bibinfo  {journal} {Phys. Rev. Lett.}\ }\textbf {\bibinfo
  {volume} {134}},\ \bibinfo {pages} {166701} (\bibinfo {year}
  {2025})}\BibitemShut {NoStop}%
\bibitem [{\citenamefont {Fu}\ and\ \citenamefont {Kane}(2008)}]{fu2008}%
  \BibitemOpen
  \bibfield  {author} {\bibinfo {author} {\bibfnamefont {Liang}\ \bibnamefont
  {Fu}}\ and\ \bibinfo {author} {\bibfnamefont {C.~L.}\ \bibnamefont {Kane}},\
  }\bibfield  {title} {\enquote {\bibinfo {title} {Superconducting proximity
  effect and majorana fermions at the surface of a topological insulator},}\
  }\href {\doibase 10.1103/PhysRevLett.100.096407} {\bibfield  {journal}
  {\bibinfo  {journal} {Phys. Rev. Lett.}\ }\textbf {\bibinfo {volume} {100}},\
  \bibinfo {pages} {096407} (\bibinfo {year} {2008})}\BibitemShut {NoStop}%
\bibitem [{\citenamefont {Kallin}\ and\ \citenamefont
  {Berlinsky}(2016)}]{Kallin2016}%
  \BibitemOpen
  \bibfield  {author} {\bibinfo {author} {\bibfnamefont {Catherine}\
  \bibnamefont {Kallin}}\ and\ \bibinfo {author} {\bibfnamefont {John}\
  \bibnamefont {Berlinsky}},\ }\bibfield  {title} {\enquote {\bibinfo {title}
  {Chiral superconductors},}\ }\href {\doibase 10.1088/0034-4885/79/5/054502}
  {\bibfield  {journal} {\bibinfo  {journal} {Reports on Progress in Physics}\
  }\textbf {\bibinfo {volume} {79}},\ \bibinfo {pages} {054502} (\bibinfo
  {year} {2016})}\BibitemShut {NoStop}%
\bibitem [{\citenamefont {Gonz\'alez-Hern\'andez}\ \emph
  {et~al.}(2021)\citenamefont {Gonz\'alez-Hern\'andez}, \citenamefont
  {\ifmmode~\check{S}\else \v{S}\fi{}mejkal}, \citenamefont {V\'yborn\'y},
  \citenamefont {Yahagi}, \citenamefont {Sinova}, \citenamefont {Jungwirth},\
  and\ \citenamefont {\ifmmode~\check{Z}\else
  \v{Z}\fi{}elezn\'y}}]{Zelezny2021}%
  \BibitemOpen
  \bibfield  {author} {\bibinfo {author} {\bibfnamefont {Rafael}\ \bibnamefont
  {Gonz\'alez-Hern\'andez}}, \bibinfo {author} {\bibfnamefont {Libor}\
  \bibnamefont {\ifmmode~\check{S}\else \v{S}\fi{}mejkal}}, \bibinfo {author}
  {\bibfnamefont {Karel}\ \bibnamefont {V\'yborn\'y}}, \bibinfo {author}
  {\bibfnamefont {Yuta}\ \bibnamefont {Yahagi}}, \bibinfo {author}
  {\bibfnamefont {Jairo}\ \bibnamefont {Sinova}}, \bibinfo {author}
  {\bibfnamefont {Tom\'a\ifmmode \check{s}\else~\v{s}\fi{}}\ \bibnamefont
  {Jungwirth}}, \ and\ \bibinfo {author} {\bibfnamefont {Jakub}\ \bibnamefont
  {\ifmmode~\check{Z}\else \v{Z}\fi{}elezn\'y}},\ }\bibfield  {title} {\enquote
  {\bibinfo {title} {Efficient electrical spin splitter based on
  nonrelativistic collinear antiferromagnetism},}\ }\href {\doibase
  10.1103/PhysRevLett.126.127701} {\bibfield  {journal} {\bibinfo  {journal}
  {Phys. Rev. Lett.}\ }\textbf {\bibinfo {volume} {126}},\ \bibinfo {pages}
  {127701} (\bibinfo {year} {2021})}\BibitemShut {NoStop}%
\bibitem [{\citenamefont {McClarty}\ and\ \citenamefont {Rau}(2024)}]{strain}%
  \BibitemOpen
  \bibfield  {author} {\bibinfo {author} {\bibfnamefont {Paul~A.}\ \bibnamefont
  {McClarty}}\ and\ \bibinfo {author} {\bibfnamefont {Jeffrey~G.}\ \bibnamefont
  {Rau}},\ }\bibfield  {title} {\enquote {\bibinfo {title} {Landau theory of
  altermagnetism},}\ }\href {\doibase 10.1103/PhysRevLett.132.176702}
  {\bibfield  {journal} {\bibinfo  {journal} {Phys. Rev. Lett.}\ }\textbf
  {\bibinfo {volume} {132}},\ \bibinfo {pages} {176702} (\bibinfo {year}
  {2024})}\BibitemShut {NoStop}%
\bibitem [{\citenamefont {Amin}\ \emph {et~al.}(2024)\citenamefont {Amin},
  \citenamefont {Dal~Din}, \citenamefont {Golias}, \citenamefont {Niu},
  \citenamefont {Zakharov}, \citenamefont {Fromage}, \citenamefont {Fields},
  \citenamefont {Heywood}, \citenamefont {Cousins}, \citenamefont
  {Maccherozzi}, \citenamefont {Krempask{\'y}}, \citenamefont {Dil},
  \citenamefont {Kriegner}, \citenamefont {Kiraly}, \citenamefont {Campion},
  \citenamefont {Rushforth}, \citenamefont {Edmonds}, \citenamefont {Dhesi},
  \citenamefont {{\v S}mejkal}, \citenamefont {Jungwirth},\ and\ \citenamefont
  {Wadley}}]{Amin2024}%
  \BibitemOpen
  \bibfield  {author} {\bibinfo {author} {\bibfnamefont {O.~J.}\ \bibnamefont
  {Amin}}, \bibinfo {author} {\bibfnamefont {A.}~\bibnamefont {Dal~Din}},
  \bibinfo {author} {\bibfnamefont {E.}~\bibnamefont {Golias}}, \bibinfo
  {author} {\bibfnamefont {Y.}~\bibnamefont {Niu}}, \bibinfo {author}
  {\bibfnamefont {A.}~\bibnamefont {Zakharov}}, \bibinfo {author}
  {\bibfnamefont {S.~C.}\ \bibnamefont {Fromage}}, \bibinfo {author}
  {\bibfnamefont {C.~J.~B.}\ \bibnamefont {Fields}}, \bibinfo {author}
  {\bibfnamefont {S.~L.}\ \bibnamefont {Heywood}}, \bibinfo {author}
  {\bibfnamefont {R.~B.}\ \bibnamefont {Cousins}}, \bibinfo {author}
  {\bibfnamefont {F.}~\bibnamefont {Maccherozzi}}, \bibinfo {author}
  {\bibfnamefont {J.}~\bibnamefont {Krempask{\'y}}}, \bibinfo {author}
  {\bibfnamefont {J.~H.}\ \bibnamefont {Dil}}, \bibinfo {author} {\bibfnamefont
  {D.}~\bibnamefont {Kriegner}}, \bibinfo {author} {\bibfnamefont
  {B.}~\bibnamefont {Kiraly}}, \bibinfo {author} {\bibfnamefont {R.~P.}\
  \bibnamefont {Campion}}, \bibinfo {author} {\bibfnamefont {A.~W.}\
  \bibnamefont {Rushforth}}, \bibinfo {author} {\bibfnamefont {K.~W.}\
  \bibnamefont {Edmonds}}, \bibinfo {author} {\bibfnamefont {S.~S.}\
  \bibnamefont {Dhesi}}, \bibinfo {author} {\bibfnamefont {L.}~\bibnamefont
  {{\v S}mejkal}}, \bibinfo {author} {\bibfnamefont {T.}~\bibnamefont
  {Jungwirth}}, \ and\ \bibinfo {author} {\bibfnamefont {P.}~\bibnamefont
  {Wadley}},\ }\bibfield  {title} {\enquote {\bibinfo {title} {Nanoscale
  imaging and control of altermagnetism in mnte},}\ }\href {\doibase
  10.1038/s41586-024-08234-x} {\bibfield  {journal} {\bibinfo  {journal}
  {Nature}\ }\textbf {\bibinfo {volume} {636}},\ \bibinfo {pages} {348--353}
  (\bibinfo {year} {2024})}\BibitemShut {NoStop}%
\bibitem [{\citenamefont {Tinkham}(2004)}]{tinkham2004}%
  \BibitemOpen
  \bibfield  {author} {\bibinfo {author} {\bibfnamefont {Michael}\ \bibnamefont
  {Tinkham}},\ }\href {https://store.doverpublications.com/0486435032.html}
  {\emph {\bibinfo {title} {Introduction to superconductivity}}}\ (\bibinfo
  {publisher} {Dover Publications},\ \bibinfo {year} {2004})\BibitemShut
  {NoStop}%
\bibitem [{\citenamefont {Rosa}\ and\ \citenamefont {Cohen}(1907)}]{Cohen1907}%
  \BibitemOpen
  \bibfield  {author} {\bibinfo {author} {\bibfnamefont {E.~B.}\ \bibnamefont
  {Rosa}}\ and\ \bibinfo {author} {\bibfnamefont {L.}~\bibnamefont {Cohen}},\
  }\bibfield  {title} {\enquote {\bibinfo {title} {On the self-inductance of
  circles},}\ }\href {\doibase 10.6028/bulletin.083} {\bibfield  {journal}
  {\bibinfo  {journal} {Bulletin of the Bureau of Standards}\ }\textbf
  {\bibinfo {volume} {4}},\ \bibinfo {pages} {149} (\bibinfo {year}
  {1907})}\BibitemShut {NoStop}%
\bibitem [{\citenamefont {Dai}(2015)}]{Dai2015}%
  \BibitemOpen
  \bibfield  {author} {\bibinfo {author} {\bibfnamefont {Pengcheng}\
  \bibnamefont {Dai}},\ }\bibfield  {title} {\enquote {\bibinfo {title}
  {Antiferromagnetic order and spin dynamics in iron-based superconductors},}\
  }\href {\doibase 10.1103/RevModPhys.87.855} {\bibfield  {journal} {\bibinfo
  {journal} {Rev. Mod. Phys.}\ }\textbf {\bibinfo {volume} {87}},\ \bibinfo
  {pages} {855--896} (\bibinfo {year} {2015})}\BibitemShut {NoStop}%
\bibitem [{\citenamefont {Pfleiderer}(2009)}]{Pfleiderer2009}%
  \BibitemOpen
  \bibfield  {author} {\bibinfo {author} {\bibfnamefont {Christian}\
  \bibnamefont {Pfleiderer}},\ }\bibfield  {title} {\enquote {\bibinfo {title}
  {Superconducting phases of $f$-electron compounds},}\ }\href {\doibase
  10.1103/RevModPhys.81.1551} {\bibfield  {journal} {\bibinfo  {journal} {Rev.
  Mod. Phys.}\ }\textbf {\bibinfo {volume} {81}},\ \bibinfo {pages}
  {1551--1624} (\bibinfo {year} {2009})}\BibitemShut {NoStop}%
\bibitem [{\citenamefont {Aoki}\ \emph {et~al.}(2019)\citenamefont {Aoki},
  \citenamefont {Ishida},\ and\ \citenamefont {Flouquet}}]{Aoki2019}%
  \BibitemOpen
  \bibfield  {author} {\bibinfo {author} {\bibfnamefont {Dai}\ \bibnamefont
  {Aoki}}, \bibinfo {author} {\bibfnamefont {Kenji}\ \bibnamefont {Ishida}}, \
  and\ \bibinfo {author} {\bibfnamefont {Jacques}\ \bibnamefont {Flouquet}},\
  }\bibfield  {title} {\enquote {\bibinfo {title} {Review of u-based
  ferromagnetic superconductors: Comparison between uge2, urhge, and ucoge},}\
  }\href {\doibase 10.7566/JPSJ.88.022001} {\bibfield  {journal} {\bibinfo
  {journal} {Journal of the Physical Society of Japan}\ }\textbf {\bibinfo
  {volume} {88}},\ \bibinfo {pages} {022001} (\bibinfo {year}
  {2019})}\BibitemShut {NoStop}%
\bibitem [{\citenamefont {Chakraborty}\ and\ \citenamefont
  {Black-Schaffer}(2024)}]{Chakraborty2024}%
  \BibitemOpen
  \bibfield  {author} {\bibinfo {author} {\bibfnamefont {Debmalya}\
  \bibnamefont {Chakraborty}}\ and\ \bibinfo {author} {\bibfnamefont
  {Annica~M.}\ \bibnamefont {Black-Schaffer}},\ }\href
  {https://arxiv.org/abs/2309.14427} {\enquote {\bibinfo {title} {Zero-field
  finite-momentum and field-induced superconductivity in altermagnets},}\ }
  (\bibinfo {year} {2024}),\ \Eprint {http://arxiv.org/abs/2309.14427}
  {arXiv:2309.14427 [cond-mat.supr-con]} \BibitemShut {NoStop}%
\bibitem [{\citenamefont {Sheehy}\ \emph {et~al.}(2004)\citenamefont {Sheehy},
  \citenamefont {Davis},\ and\ \citenamefont {Franz}}]{Sheehy2004}%
  \BibitemOpen
  \bibfield  {author} {\bibinfo {author} {\bibfnamefont {Daniel~E.}\
  \bibnamefont {Sheehy}}, \bibinfo {author} {\bibfnamefont {T.~P.}\
  \bibnamefont {Davis}}, \ and\ \bibinfo {author} {\bibfnamefont
  {M.}~\bibnamefont {Franz}},\ }\bibfield  {title} {\enquote {\bibinfo {title}
  {Unified theory of the $\mathrm{ab}$-plane and c-axis penetration depths of
  underdoped cuprates},}\ }\href {\doibase 10.1103/PhysRevB.70.054510}
  {\bibfield  {journal} {\bibinfo  {journal} {Phys. Rev. B}\ }\textbf {\bibinfo
  {volume} {70}},\ \bibinfo {pages} {054510} (\bibinfo {year}
  {2004})}\BibitemShut {NoStop}%
\bibitem [{\citenamefont {Tummuru}\ \emph {et~al.}(2022)\citenamefont
  {Tummuru}, \citenamefont {Plugge},\ and\ \citenamefont
  {Franz}}]{Tummuru2022}%
  \BibitemOpen
  \bibfield  {author} {\bibinfo {author} {\bibfnamefont {Tarun}\ \bibnamefont
  {Tummuru}}, \bibinfo {author} {\bibfnamefont {Stephan}\ \bibnamefont
  {Plugge}}, \ and\ \bibinfo {author} {\bibfnamefont {Marcel}\ \bibnamefont
  {Franz}},\ }\bibfield  {title} {\enquote {\bibinfo {title} {Josephson effects
  in twisted cuprate bilayers},}\ }\href {\doibase 10.1103/PhysRevB.105.064501}
  {\bibfield  {journal} {\bibinfo  {journal} {Phys. Rev. B}\ }\textbf {\bibinfo
  {volume} {105}},\ \bibinfo {pages} {064501} (\bibinfo {year}
  {2022})}\BibitemShut {NoStop}%
\end{thebibliography}%

\end{document}